\documentclass[aps,superscriptaddress,amsmath,amssymb,floatfix,twocolumn,showpacs,amsfonts,longbibliography]{revtex4-2}
\usepackage{times}
\usepackage[varg]{txfonts}
\usepackage{textcomp}
\usepackage{graphicx}
\usepackage{subfigure}
\usepackage{tabu}
\usepackage{color}
\usepackage[colorlinks=true,citecolor=blue,urlcolor=blue,linkcolor=blue,hyperindex]{hyperref}
\usepackage{braket}
\usepackage{float}
\usepackage{overpic}
\usepackage{amssymb}
\usepackage{amsmath}
\usepackage{gensymb}
\usepackage{mhchem}
\usepackage{mathrsfs}
\usepackage{bm}

\allowdisplaybreaks

\begin{document}

\title{Constraints from global symmetries of quantum ice on the planar pyrochlore lattice}
\author{Zijian Xiong}
\affiliation{College of Physics and Electronic Engineering, Chongqing Normal University, Chongqing 401331, China}

\begin{abstract}
We study the low energy effective theory of spin-1/2 quantum ice on the planar pyrochlore lattice from the perspective of its global symmetries. We show that the Fourier transform of the spin operator, $S^{z}_{\textbf{Q}}$, can be expressed in terms of conserved quantities associated with a 1-form U(1) symmetry along two high-symmetry momentum paths, leading to constraints on both the static and dynamical structure factors. We also derive Lieb-Schultz-Mattis type constraints that forbid the existence of a unique gapped ground state in the low energy effective theory. Our findings also apply to the (2+1) dimensional U(1) quantum link model.
\end{abstract}

\maketitle

\section{Introduction}
Global symmetry provides a foundational framework in physics, serving as an indispensable tool for characterizing phases of matter and understanding the nature of phase transitions. In recent years, this concept has been substantially extended in multiple directions, leading to the development of generalized notions of symmetry \cite{Gaiotto2015,McGreevy2023,Luo2024,Bhardwaj2024} such as higher-form and non-invertible symmetries \cite{shao2023,Nameki2024}. In certain cases, specifying the global symmetries is sufficient to constrain the possible ground states of a many-body system. A celebrated example is the Lieb-Schultz-Mattis (LSM) theorem \cite{LSM1961} and its subsequent generalizations \cite{Zou2026,Oshikawa2000prl,Hastings2004prb,Haruki2015pnas,YuanYao2020prx,HCPo2017prl}. Recently, it has been recognized that the LSM theorem can be understood as an anomaly \cite{Zou2026,MengCheng2016prx,Cho2017prb,Metlitski2018prb,MengCheng2022,Sahand2024} that forbids a symmetric trivial gapped ground state.

 In condensed matter physics, the low energy effective theories of certain systems are well described by emergent gauge theories, which may possess higher-form symmetries \cite{XGWen2019prb,CMJian2021,Pace2023prb,Kobayashi2019prb}. Notable examples include quantum dimer models \cite{Fradkin2013book}, quantum spin liquids \cite{Savary2017}, and quantum spin ice \cite{Fradkin2006prb,Gingras2014}. In classical spin ice on the three dimensional pyrochlore lattice \cite{Anderson1956pr}, the local "two-in-two-out" ice rule gives rise to a macroscopically degenerate manifold and a divergence-free effective gauge field \cite{Gingras2014}. When quantum fluctuations are perturbatively introduced, tunneling among the degenerate classical configurations yields an emergent quantum electrodynamics description \cite{Hermele2004prb} that has been extensively studied both theoretically and experimentally (see ref.\cite{Moessner2012spinice,Gingras2014,udagawa2021book} for recent reviews). 
 
The square ice on the planar pyrochlore lattice serves as a two dimensional analog of the pyrochlore classical spin ice. The Ising model on this lattice can be mapped to the six vertex model, which was solved exactly by Lieb \cite{Lieb1967pr}, revealing a massively degenerate ground state and algebraically decaying correlations. When quantum dynamics is introduced into this classical degenerate ground state manifold either via transverse exchange interactions \cite{Shannon2004prb} or a transverse field \cite{Moessner2001prb,Henry2014prl,Langari2019prb}, the classical degeneracy is lifted by quantum fluctuations through the so-called order by disorder mechanism \cite{Moessner2001prb,Fradkin2006prb}, and the resulting quantum ground state is a plaquette valence bond solid \cite{Shannon2004prb,Henry2014prl,Langari2019prb}. Remarkably, both types of quantum fluctuations give rise, at lowest order non-trivial contributions, to the same four-spin ring exchange term under degenerate perturbation theory \cite{Fradkin2006prb}. Combined with the ice rule constraint, the low energy effective theory of the quantum square ice is described by a pure U(1) lattice gauge theory without matter fields \cite{Fradkin2006prb,Henry2014prl}. Further adding a Rokhsar-Kivelson (RK) potential lead to the RK model \cite{Kivelson1988prl}, which is equivalent to the quantum link model \cite{Banerjee2013Jstat,Tschirsich2019scipost,Xiaoxue2022}. The phase diagram of the latter has also been extensively studied \cite{Banerjee2013Jstat,Tschirsich2019scipost,Xiaoxue2022}, where all phases are found to be either symmetry broken or gapless.
 
The order by disorder phenomenon in quantum square ice can be deduced via duality transformation \cite{Fradkin2006prb} or height mapping \cite{Moessner2001prb} within the low energy effective theory. Alternatively, some information about the ground state may be inferred from LSM type arguments applied to either the microscopic lattice model or the low energy theory. LSM type constraints can be broadly classified into three categories \cite{Tasaki2022review,Sahand2024,Zou2026}. The first originates from the original LSM theorem \cite{LSM1961}, which relies on a variational argument and a twist operator. The second applies to systems with U(1) symmetry and is based on the U(1) flux insertion argument \cite{Oshikawa2000prl}, which forbids a symmetric trivial ground state when the system has a fractional filling factor \cite{Oshikawa2000prl,Oshikawa1997prl}. The third type relies on the nontrivial projective representation of the symmetry \cite{YuanYao2021prl,HCPo2017prl,YuanYao2022prb,Haruki2015pnas,XChen2011prb} . The latter two approaches are commonly used in the system with dimensions higher than one and in the presence of various symmetries.
 
Two microscopic lattice models are widely used to study the physics of quantum square ice: transverse field Ising model and XXZ model on the planar pyrochlore lattice. Zeeman field can also be included. Both models have the same low energy effective theory. The transverse field Ising model has relatively low symmetry, and thus does not admit LSM constraints from the above treatments. For the XXZ model, the third type of LSM treatment \cite{HCPo2017prl,Tada2026prb} yields nontrivial constraints for the ground state, but becomes silent once the Zeeman field is introduced. Moreover, since there are two spins per unit cell in the planar pyrochlore lattice, ref.\cite{Furuya2019prb} shows that the second type of LSM treatment yields a constraints only when a tilted Klein bottle boundary condition is imposed. However, this boundary condition requires the system size to be $L\times L$ and is not applicable to other system sizes.
 
The low energy effective theory of quantum square ice is described by a pure U(1) lattice gauge theory \cite{Moessner2001prb,Shannon2004prb,Fradkin2006prb,Henry2014prl}, and there is a 1-form U(1) symmetry \cite{XGWen2019prb,CMJian2021,Pace2023prb,Kobayashi2019prb}. It has recently become clear that such higher-form symmetries can also lead to nontrivial constraints for the ground states of lattice systems \cite{Kobayashi2019prb,Pace2023prb}. Concurrently, lattice models that admit an emergent gauge theory description have attracted significant interest, driven in part by their recent experimental realizations in platforms such as Rydberg atom arrays \cite{Surace2020prx,HZhai2024,Lukin2025} and superconducting qubit systems \cite{King2020,King2023}. In this work, we investigate the global symmetries and their physical consequences for quantum ice on the planar pyrochlore lattice. We find that the 1-form U(1) symmetry gives rise to a selection rule that applies to both the static and dynamical structure factors. And we also derive LSM type constraints within the low energy effective theory that require lower spatial symmetry than that obtained from the U(1) flux insertion argument \cite{Furuya2019prb}.

 This paper is organized as follows. In section \ref{sectmod}, we introduce the low energy effective model of a spin-1/2 XXZ model on the planar pyrochlore lattice and show how it maps onto a U(1) lattice gauge theory. We also discuss the closely related U(1) quantum link model, the relations among these models, and their global symmetries. We present our main results in Sec.~\ref{sectres}, where we derive constraints imposed by the global symmetries of these models.

\section{Model}\label{sectmod}
\begin{figure}
	\centering
	\includegraphics[width=0.48\textwidth]{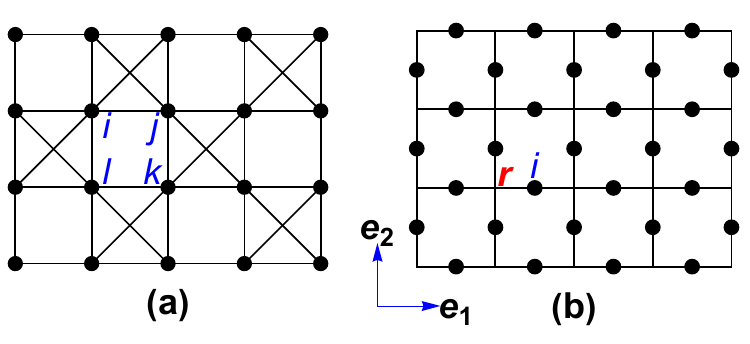}
	\caption{\label{lattice}  (a) Planar pyrochlore lattice. The black dots denote the spins. (b) A square lattice with spins located on its links. This structure is useful in the lattice gauge theory discussed in the main text.}
\end{figure}

 We consider the spin-1/2 XXZ model under a Zeeman field:
\begin{equation}\label{lattuv}
H=-J_{xy}\sum_{\langle i,j\rangle}(S^{+}_{i}S^{-}_{j}+S^{-}_{i}S^{+}_{j})+J_{z}\sum_{\langle i,j\rangle}S^{z}_{i}S^{z}_{j}-h\sum_{i}S^{z}_{i},
\end{equation}
where $\langle i,j\rangle$ denotes the nearest-neighbor bonds and connected diagonal bonds as illustrated in Fig.\ref{lattice} (a). This model is equivalent to an extended Bose-Hubbard model via the mapping: $S^{+}_i\to b^{\dagger}_i, S^{-}_i \to b_i, S^{z}_{i}\to n_i-1/2$, where $b^{\dagger}_i, b_i$ are bosonic creation and annihilation operators at site $i$, and $n_i= b^{\dagger}_i b_i$ denotes the number operator. 

 We first consider the classical part of the model,
\begin{equation}
H^{0}=J_{z}\sum_{\langle i,j\rangle}S^{z}_{i}S^{z}_{j}-h\sum_{i}S^{z}_{i},
\end{equation}
using the corner-sharing structure of the planar pyrochlore lattice, this can be rewritten as
\begin{equation}
H^{0}=\frac{J_z}{2}\sum_{\boxtimes}\left[(\sum_{i\in\boxtimes}S_{i}^{z})-\frac{h}{2J_z}\right]^2-C(h,J_z)
\end{equation} 
where $\boxtimes$ denotes a crossing plaquette with diagonal bonds. $C(h,J_z)$ is constant for fixed $h,J_z$ and will be omitted in the following discussion. The possible values for $\sum_{i\in\boxtimes}S_{i}^{z}$ are $-2,-1,0,1,2$. In this work, we consider $J_{z}>0$ and focus on the parameter regime $-\frac{1}{2}<\frac{h}{2J_z}<\frac{1}{2}$, where the energy of $H^{0}$ is minimized by the ice rule \cite{Lieb1967pr}: $\sum_{i\in\boxtimes}S_{i}^{z}=0$ for every crossing plaquette. In other words, classical ground states consist of exactly two up spins and two down spins in each crossing plaquette. These states form extensively degenerate ice manifold \cite{Lieb1967pr}.  

Around the classical Ising limit, where $|J_{xy}|\ll J_z$, small quantum fluctuations induce quantum tunneling within the ice manifold. The effective Hamiltonian within the ice manifold can be derived using degenerate perturbation theory \cite{Kato1949,Klein1974,Takahashi1977} (also see appendix \ref{appendix}). To the lowest order \cite{Shannon2004prb,Fradkin2006prb,Henry2014prl}, one obtains ring-exchange terms
\begin{equation}\label{ringex}
H^{\rm{eff}}=-t\sum_{\square}(S^{+}_{i}S^{-}_{j}S^{+}_{k}S^{-}_{l}+S^{-}_{i}S^{+}_{j}S^{-}_{k}S^{+}_{l}),
\end{equation}
 where $\square$ denotes a plaquette without diagonal bonds, and $i,j,k,l$ label the four spins around such a plaquette in cyclic order (see Fig.\ref{lattice} (a)). And $t=4J_{xy}^2/J_{z}$. Notably, the lowest order effective theory for the transverse field Ising model \cite{Moessner2001prb,Fradkin2006prb,Henry2014prl} around the Ising limit has the same form as eq.(\ref{ringex}). 
 
 To restrict the dynamics to the ice manifold, we can enforce the ice rule energetically \cite{Cenke2016,Hastings2005prb,Pace2023prb}
 \begin{equation}\label{ringice}
H^{\rm{eff}}_{\rm{ice}}=-t\sum_{\square}(S^{+}_{i}S^{-}_{j}S^{+}_{k}S^{-}_{l}+S^{-}_{i}S^{+}_{j}S^{-}_{k}S^{+}_{l})+\frac{J_z}{2}\sum_{\boxtimes}(\sum_{i\in\boxtimes}S_{i}^{z})^2,
\end{equation}
with $J_z\gg|t|$. A key feature is that each $\sum_{i\in\boxtimes}S_{i}^{z}\equiv S^{z}_{\boxtimes}$ commutes with all other terms in $H^{\rm{eff}}_{\rm{ice}}$. Thus, there are local conserved quantities $S^{z}_{\boxtimes}$ and an associated local U(1) symmetry \cite{Hermele2004prb} generated by $U_{\boxtimes}(\theta)=\rm{exp}(i\theta S^{z}_{\boxtimes})$ in the effective model $H^{\rm{eff}}_{\rm{ice}}$. More generally, to all orders in degenerate perturbation theory, every term in the effective theory must commute with $S^{z}_{\boxtimes}$, otherwise, it will violate the ice rule. This implies the high-order terms in the effective theory necessarily take the form of loop operators.

\subsection{Lattice gauge theory}
 To study the local symmetry of the model in eq. (\ref{ringice}), it is valuable to map it to a lattice gauge theory \cite{Hermele2004prb,Fradkin2006prb}. In this framework, the spin operators are conveniently expressed in terms of link variables. We introduce a square lattice with spins located on the links. For example, $S^{z}_{i}=S^{z}_{1}(\textbf{r})$ in Fig.\ref{lattice} (b), where $\textbf{r}$ denotes the vertex of the square lattice and spin locate at the link $(\textbf{r},\textbf{r}+\textbf{e}_1)$. Generally, the spin operator is denoted by $S^{\alpha}_{\mu}(\mathbf{r})$. Here, the index $\alpha = x, y, z$ specifies the spin component, and $\mu = 1, 2$ refers to the lattice directions $\mathbf{e}_1$ and $\mathbf{e}_2$. Similar to the toric code model \cite{kitaev2003}, the $t$ term and $J_z$ term in eq.(\ref{ringice}) correspond to operators defined on the plaquettes and vertices of the square lattice, respectively. Therefore, we will focus on model eq.(\ref{ringice}) defined on the square lattice. Now, the ice rule term $\sum_{\boxtimes}(\sum_{i\in\boxtimes}S_{i}^{z})^2$ can be rewritten as $\sum_{\textbf{r}}(\sum_{i\in\textbf{r}}S_{i}^{z})^2$.

We now introduce a rotor representation for the spins \cite{Hermele2004prb}: $ S^{z}_{\mu}(\textbf{r})=n_{\mu}(\textbf{r})-1/2$, $S^{\pm}_{\mu}(\textbf{r})=e^{\pm i\phi_{\mu}(\textbf{r})}$, and the canonical commutation relation reads $[\phi_{\mu}(\textbf{r}),n_{\nu}(\textbf{r}')]=i\delta_{\mu,\nu}\delta_{\textbf{r},\textbf{r}'}$. For spin-1/2, the eigenvalue of $n_{\mu}(\textbf{r})$ is either 0 or 1, corresponding to spin up and down, respectively. And, eigenvalues of $\phi_{\mu}(\textbf{r})$ take value in $[0,2\pi)$. One can verify that this representation reproduces the spin commutation relation $[S^{\pm},S^{z}]=\mp S^{\pm}$. 

 To strictly enforce the spin-$1/2$ algebra, one should also introduce a projection \cite{Benton2012prb} (which projects to the physical space where $n_i$ takes $0,1$) in the rotor representation of the ladder operators: $S^{+}_i=\sqrt{n_i} e^{i \phi_i} \sqrt{1-n_i}$, $S^{-}_i=\sqrt{1-n_i}e^{-i\phi_i}\sqrt{n_i}$ to keep $[S^{+},S^{-}]=2S^{z}$ invariant. An equivalent and more practical approach is to add an energetic penalty term \cite{Benton2012prb,Hermele2004prb,Fradkin2006prb} $\frac{U}{2}\sum_i[(n_i-\frac{1}{2})^2-\frac{1}{4}]$ and take the limit $U\to \infty$ to project onto the physical subspace. Applying this rotor representation to the effective model in eq. (\ref{ringice}) yields
\begin{equation}
\begin{aligned}\label{rotor}
H^{\rm{eff}}_{\rm{rot}}=&-2t\sum_{\textbf{r}}\cos[\phi_{1}(\textbf{r})-\phi_2(\textbf{r}+\textbf{e}_1)+\phi_1(\textbf{r}+\textbf{e}_2)-\phi_2(\textbf{r})]\\
&+\frac{J_z}{2}\sum_{\textbf{r}}\left[\sum_{i\in\textbf{r}}(n_i-\frac{1}{2})\right]^2+\frac{U}{2}\sum_i[(n_i-\frac{1}{2})^2-\frac{1}{4}].
\end{aligned}
\end{equation}

 Since the square lattice is bipartite, we can define an orientation on its bonds. For instance, we can take the positive orientation to point from the A sublattice to the B sublattice. This allows us to introduce oriented link variables \cite{Hermele2004prb}: $E_\mu(\textbf{r})=\pm(n_\mu(\textbf{r})-\frac{1}{2})$, $a_\mu(\textbf{r})=\pm\phi_\mu(\textbf{r})$, where $+$ ($-$) sign applies when $\textbf{r}$ belongs to the A (B) sublattice. These link variables behave as vectors, for instance, $E_{-\mu}(\textbf{r}+\textbf{e}_\mu)=-E_{\mu}(\textbf{r})$. The vertices of the square lattice are labeled by vectors $\textbf{r}=m \textbf{e}_1+n \textbf{e}_2$ where $m,n$ are integers and we set the lattice constant to 1. Further, we assign sites with even $m+n$ to the A sublattice, and sites with odd $m+n$ to the B sublattice. The sign factor can then be absorbed into a phase \cite{Fradkin2013book} as the square lattice is a Bravais lattice, 
\begin{equation}
\begin{aligned}\label{stagg}
E_{\mu}(\textbf{r})=&e^{i\textbf{Q}_{0}\cdot \textbf{r}}(n_{\mu}(\textbf{r})-\frac{1}{2}),\\
a_{\mu}(\textbf{r})=&e^{i\textbf{Q}_{0}\cdot \textbf{r}}\phi_{\mu}(\textbf{r}),
\end{aligned}
\end{equation}
where $\textbf{Q}_{0}=(\pi,\pi)$. These variables satisfy the canonical commutation relation $[a_{\mu}(\textbf{r}), E_{\nu}(\textbf{r}')]=i\delta_{\mu\nu}\delta_{\textbf{r},\textbf{r}'}$ where $\mu,\nu=1,2$. Finally, we note that $E_{\mu}(\textbf{r})=e^{i\textbf{Q}_{0}\cdot \textbf{r}}S^{z}_{\mu}(\textbf{r})$ maps the spin model in eq.(\ref{ringex}) to the six vertex model \cite{Shannon2004prb}.

We now reformulate the effective model as a lattice gauge theory \cite{Fradkin2013book}. 

i). Gauss law. Consider the ice rule term,
\begin{equation}
\begin{aligned}\label{lattdiv}
\sum_{i\in\textbf{r}}S_{i}^{z}=&S^{z}_{1}(\textbf{r})+S^{z}_{2}(\textbf{r})+S^{z}_{1}(\textbf{r}-\textbf{e}_{1})+S^{z}_{2}(\textbf{r}-\textbf{e}_{2})\\
=&e^{i\textbf{Q}_{0}\cdot \textbf{r}}\left[E^{z}_{1}(\textbf{r})+E^{z}_{2}(\textbf{r})-E^{z}_{1}(\textbf{r}-\textbf{e}_1)-E^{z}_{2}(\textbf{r}-\textbf{e}_2)\right]\\
=&e^{i\textbf{Q}_{0}\cdot \textbf{r}}\sum_{\mu=1}^2 \Delta^{-}_\mu E_\mu(\textbf{r})\equiv e^{i\textbf{Q}_{0}\cdot \textbf{r}}(\rm{div}  \,\mathnormal{E})_{\textbf{r}},
\end{aligned}
\end{equation}
where we have introduced the lattice derivative \cite{Ardonne2004} $\Delta^{-}_{\mu}f(\textbf{r})=f(\textbf{r})-f(\textbf{r}-\textbf{e}_{\mu})$ and defined the lattice divergence: $(\rm{div}\,\mathnormal{E})_{\textbf{r}}$. Therefore, the ice rule translates into a Gauss law in the lattice gauge theory. Here, the Gauss law is imposed energetically \cite{Cenke2016,MengCheng2019prb,Choi2025} in the Hamiltonian. 

ii). Curl. Consider $t$ term in eq.(\ref{rotor}),
\begin{equation}
\begin{aligned}
&\phi_{1}(\textbf{r})-\phi_2(\textbf{r}+\textbf{e}_1)+\phi_1(\textbf{r}+\textbf{e}_2)-\phi_2(\textbf{r})\\
=&e^{i\textbf{Q}_{0}\cdot \textbf{r}}\left[a_{1}(\textbf{r})+a_2(\textbf{r}+\textbf{e}_1)-a_1(\textbf{r}+\textbf{e}_2)-a_2(\textbf{r})\right]\\
=&e^{i\textbf{Q}_{0}\cdot \textbf{r}}\left[\Delta_1^{+}a_{2}(\textbf{r})-\Delta^{+}_{2}a_{1}(\textbf{r})\right]\\
=&e^{i\textbf{Q}_{0}\cdot \textbf{r}}\sum_{\mu,\nu=1}^{2}\epsilon_{\mu\nu}\Delta^{+}_{\mu}a_\nu(\textbf{r})\equiv e^{i\textbf{Q}_{0}\cdot \textbf{r}}  (\rm{curl}\, \textit{a})_\textbf{r},
\end{aligned}
\end{equation}
where we have used the lattice derivative \cite{Ardonne2004} $\Delta^{+}_{\mu}f(\textbf{r})=f(\textbf{r}+\textbf{e}_\mu)-f(\textbf{r})$ and Levi-Civita symbol: $\epsilon_{1,2}=-\epsilon_{2,1}=1$ and 0 if any index is repeated. And we have also defined the lattice curl:  $(\rm{curl}\, \textit{a})_\textbf{r}$.

iii). Lattice gauge theory. With these settings, the effective model eq.(\ref{rotor}) becomes 
\begin{equation}\label{lgt}
\begin{aligned}
 H^{\rm{LGT}}=&-2t\,\sum_{\textbf{r}}\cos[(\rm{curl}\,\textit{a})_{\textbf{r}}] +\frac{\mathnormal{U}}{2}\sum_{\textbf{r},\mu}[E^{2}_{\mu}(\mathbf{r})-\frac{1}{4}]\\
 &+\frac{\mathnormal{J_z}}{2}\sum_{\textbf{r}}\left[(\rm{div}\,\mathnormal{E})_{\textbf{r}}\right]^2,
 \end{aligned} 
 \end{equation}
which has the same form as a U(1) lattice gauge theory \cite{Benton2012prb,Hermele2004prb,Fradkin2006prb,Fradkin2013book}. It is not difficult to find that $[(\rm{div}\,\mathnormal{E})_{\mathbf{r}}, (\rm{curl}\,\textit{a})_{\mathbf{r}'}]=0$, hence the Gauss law term commutes with all other terms. The U(1) gauge symmetry is generated by $U=e^{i\sum_{\mathbf{r}}\theta_{\mathbf{r}}(\rm{div}\,\mathnormal{E})_{\mathbf{r}}}$, where $\theta_{\mathbf{r}}$ is c-number. One can find that $a_{\mu}(\mathbf{r})$ transform as the usual vector potential 
 \begin{equation}
 Ua_{\mu}(\mathbf{r})U^{\dagger}=a_{\mu}(\mathbf{r})+\theta_{\mathbf{r}}-\theta_{\mathbf{r}+\mathbf{e}_{\mu}}=a_{\mu}(\mathbf{r})-\Delta^{+}_{\mu}\theta_{\mathbf{r}},
 \end{equation}
 under the gauge transformation, then $E_{\mu}(\mathbf{r})$ can be recognized as electric field from the canonical commutation relation.

Another closely related model is U(1) quantum link model \cite{Shannon2004prb,Banerjee2013Jstat,Tschirsich2019scipost} 
\begin{equation}\label{qlm}
H^{\rm{QLM}}=\sum_{\textbf{r}}(-f_{\textbf{r}}+\lambda f^{2}_{\textbf{r}}),
\end{equation}
where $f_{\textbf{r}}=S^{+}_{1}(\textbf{r})S^{+}_{2}(\textbf{r}+\textbf{e}_1)S^{-}_{1}(\textbf{r}+\textbf{e}_2)S^{-}_{2}(\textbf{r})+h.c.$ is plaquette term. The Gauss law is imposed by requiring physical states $|\psi\rangle$ to satisfy $G_{\textbf{r}}|\psi\rangle=0$ where $G_{\textbf{r}}=S^{z}_{1}(\textbf{r})+S^{z}_{2}(\textbf{r})-S^{z}_{1}(\textbf{r}-\textbf{e}_1)-S^{z}_{2}(\textbf{r}-\textbf{e}_2)$ is equivalent to the lattice divergence term in eq.(\ref{lattdiv}). One can find that $f_{\textbf{r}}$ is equivalent to $\cos[(\rm{curl}\,\textit{a})_{\textbf{r}}]$ in eq.(\ref{lgt}), and is also unitary equivalent to the $t$ term in eq.(\ref{ringex}) under a transformation applied to one sublattice $U=\prod_{\textbf{r}\in B}\rm{exp}(i\pi \sum_{\mu=1}^{2}S^{x}_{\mu}(\textbf{r}))$ (also see appendix \ref{appendixb}). This model belongs to the Rokhsar-Kivelson type model \cite{Kivelson1988prl}.

\subsection{Global symmetries}
\begin{figure}
	\centering
	\includegraphics[width=0.48\textwidth]{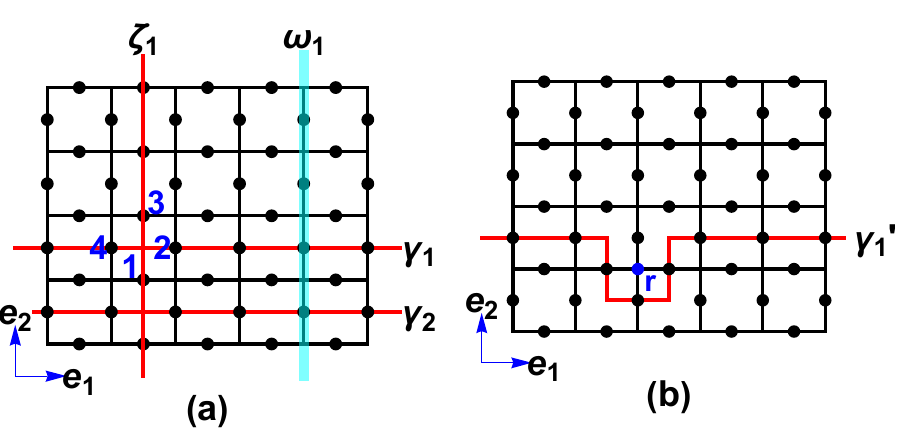}
	\caption{\label{oneform} (a) Three noncontractible loops $\gamma_1,\zeta_1,\omega_1$ of the square lattice with periodic boundary conditions. The 1-form conserved quantities $\widetilde{W}_{\gamma_1}, \widetilde{W}_{\zeta_1}$ are defined along $\gamma_1,\zeta_1$. The Wilson operator is defined along $\omega_1$. (b) $\widetilde{W}_{\gamma_1}$ can be deformed to $\widetilde{W}_{\gamma_1\,'}$ by subtracting a lattice divergence term at site $\textbf{r}$. Similarly, $\widetilde{W}_{\gamma_1}$ can be deformed to $\widetilde{W}_{\gamma_2}$.}
\end{figure}

Besides the local symmetry, the effective model eq.(\ref{ringice}), U(1) lattice gauge theory eq.(\ref{lgt}) and U(1) quantum link model eq.(\ref{qlm}) share the same global symmetries \cite{Banerjee2013Jstat,Tschirsich2019scipost,Biswas2022}:

i).Charge conjugation symmetry $C$: 
\begin{equation}
\begin{aligned}\label{charcon}
&S^{x}_i\to S^{x}_i, S^{y}_i \to -S^{y}_i, S^{z}_i\to -S^{z}_i,\\
&E_\mu(\textbf{r})\to -E_{\mu}(\textbf{r}),  a_{\mu}(\textbf{r})\to -a_{\mu}(\textbf{r}).
\end{aligned}
\end{equation}
In the spin language, this symmetry corresponds to a $\pi$ rotation about the $x$-axis, generated by $U_{x}(\pi) = \rm{exp}(-i\pi \sum_{i} S^{x}_i)$. It is important to note that this charge conjugation symmetry is emergent within the ice manifold; it is explicitly broken in the UV by the Zeeman term in the original Hamiltonian eq. (\ref{lattuv}).

ii).Lattice symmetries of square lattice (whose space group is the wallpaper group $p4m$), some of them are: a). Translation along $\textbf{e}_1,\textbf{e}_2$ direction. b). 2-fold rotation symmetry $C_2$ with rotation centers located at the middle point of the links. c). 4-fold rotation symmetry $C_4$ with rotation centers located at vertices. However, it depends on the system size: a square lattice of size $L_1\times L_2$ has this symmetry only when $L_1=L_2$, where $L_1$ and $L_2$ are the dimensions along $\textbf{e}_1,\textbf{e}_2$ directions, respectively. d). Reflection symmetry along $\textbf{e}_1,\textbf{e}_2$ direction. 

iii).1-form U(1) global symmetry. For simplicity, we begin in the spin language and then translate to the language of lattice gauge theory. Consider a plaquette term $S^{+}_1S^{-}_2S^{+}_3S^{-}_4+h.c.$, see Fig.\ref{oneform} (a), we can construct operators built from $S^{z}_i$ that commute with it. For example, $S^{z}_2-S^{z}_4$ and $S^{z}_1-S^{z}_3$. This approach inspires the construction of similar operators that commute with any given plaquette term in a system with periodic boundary conditions, two of them are \cite{Fradkin2013book,Banerjee2013Jstat,Tschirsich2019scipost,Biswas2022,Herzog2019prb}
\begin{equation}
\begin{aligned}\label{horverwinding}
\widetilde{W}_{\gamma_1}=&\sum_{\textbf{r}}^{(\gamma_1)}e^{i\textbf{Q}_{0}\cdot \textbf{r}}S^{z}_{2}(\textbf{r})=\sum_{\textbf{r}}^{(\gamma_1)}E_{2}(\textbf{r}),\\
\widetilde{W}_{\zeta_1}=&\sum_{\textbf{r}}^{(\zeta_1)}e^{i\textbf{Q}_{0}\cdot \textbf{r}}S^{z}_{1}(\textbf{r})=\sum_{\textbf{r}}^{(\zeta_1)} E_{1}(\textbf{r}),
\end{aligned}
\end{equation}
where $\gamma_1,\zeta_1$ are noncontractible loops on the torus (see Fig.\ref{oneform} (a)). We use $\gamma$ and $\zeta$ to denote the horizontal and vertical noncontractible loops, respectively. The sum $\sum_{\textbf{r}}^{(\gamma_1)}$ runs over all sites $\textbf{r}$ for which the link $(\textbf{r},\textbf{r}+\textbf{e}_2)$ intersects the loop $\gamma_1$ once, i.e., where the intersection number $\rm{Int}\left[(\textbf{r},\textbf{r}+\textbf{e}_2),\gamma_1\right]=1$.  $\sum_{\textbf{r}}^{(\zeta_1)}$ is similar, the corresponding condition is $\rm{Int}\left[(\textbf{r},\textbf{r}+\textbf{e}_1),\zeta_1\right] = 1$. Since $\widetilde{W}_{\gamma_1}$ and $\widetilde{W}_{\zeta_1}$ commute with $H^{\rm{eff}}_{\text{ice}}$ (and $H^{\rm{LGT}}$), they are conserved quantities whose eigenvalues (usually called winding numbers in the literature \cite{Kivelson1988prl,moessner2010,Herzog2019prb}) label sectors of the Hilbert space.

We can deform the loops $\gamma_1$ and $\zeta_1$ to generate other equivalent conserved quantities. For example, as illustrated in Fig.\ref{oneform} (b), $\widetilde{W}_{\gamma_1}-(\rm{div}\,\mathnormal{E})_{\textbf{r}}=\widetilde{W}_{\gamma_1\,'}$. Consequently, for eigenstates in the ice manifold, $\widetilde{W}_{\gamma_1}$ and $\widetilde{W}_{\gamma_1\,'}$ have the same eigenvalues.  We can also shift the horizontal loop by subtracting lattice divergence terms: when $L_1$ is even, we have 
\begin{equation}
\widetilde{W}_{\gamma_1}-\sum_{\textbf{r}}^{(\gamma_1)}e^{i\textbf{Q}_{0}\cdot \textbf{r}}[S^{z}_{1}(\textbf{r})+S^{z}_{2}(\textbf{r})+S^{z}_{1}(\textbf{r}-\textbf{e}_{1})+S^{z}_{2}(\textbf{r}-\textbf{e}_{2})]=\widetilde{W}_{\gamma_2},
\end{equation}
where $\gamma_2$ is shown in Fig.\ref{oneform} (a). This implies that, for eigenstates in the ice manifold, the horizontal winding numbers of every row are the same. Similarly, when $L_2$ is even, the vertical winding number of every column are the same for these eigenstates.

Based on the canonical commutation relation, two useful relations can be derived 
\begin{equation}
\begin{aligned}\label{usere}
\left[e^{i\omega a_{\mu}(\mathbf{r})},E_{\mu}(\mathbf{r}) \right]=&-\omega e^{i\omega a_{\mu}(\mathbf{r})},\\
\left[a_{\mu}(\mathbf{r}),e^{i\omega E_{\mu}(\mathbf{r})} \right]=&-\omega e^{i\omega E_{\mu}(\mathbf{r})},
\end{aligned}
\end{equation}
where $\omega$ is c-number. Finally, we consider the Wilson loop operator defined on a noncontractible loop, for example, see Fig.\ref{oneform} (a), 
\begin{equation}
W_{\omega_1}=e^{i\sum_{\textbf{r}\in \omega_1} a_{2}(\textbf{r})},
\end{equation}
using eq.(\ref{usere}), one can find 
 \begin{equation}
 \begin{aligned}\label{wcom}
 \left[W_{\omega_1}, \widetilde{W}_{\gamma_1}\right]=&-W_{\omega_1},\\
 \left[W^{\dagger}_{\omega_1}, \widetilde{W}_{\gamma_1}\right]=&W^{\dagger}_{\omega_1}.
 \end{aligned}
 \end{equation}
Consider the eigenvalue equation $\widetilde{W}_{\gamma_1}|\widetilde{w}_{\gamma_1}\rangle=\widetilde{w}_{\gamma_1}|\widetilde{w}_{\gamma_1}\rangle$, one finds $\widetilde{W}_{\gamma_1}W_{\omega_1}|\widetilde{w}_{\gamma_1}\rangle=(\widetilde{w}_{\gamma_1}+1)W_{\omega_1}|\widetilde{w}_{\gamma_1}\rangle$ and $\widetilde{W}_{\gamma_1}W^{\dagger}_{\omega_1}|\widetilde{w}_{\gamma_1}\rangle=(\widetilde{w}_{\gamma_1}-1)W^{\dagger}_{\omega_1}|\widetilde{w}_{\gamma_1}\rangle$. Hence, the Wilson operators $W_{\omega_1}, W^{\dagger}_{\omega_1}$ are  ladder operators for $\widetilde{W}_{\gamma_1}$ loop operator. Further, we have
 \begin{equation}
 e^{i\omega\widetilde{W}_{\gamma_1}}W_{\omega_1}e^{-i\omega\widetilde{W}_{\gamma_1}}=e^{i\omega}W_{\omega_1},
 \end{equation}
this imply Wilson operator $W_{\omega_1}$ (1-dimensional) is charged under a 1-form U(1) symmetry \cite{Gaiotto2015,McGreevy2023,Kobayashi2019prb,Pace2023prb} whose symmetry operator $e^{i\omega\widetilde{W}_{\gamma_1}}$ has 1-dimensional support. While the ordinary 0-form symmetry, for example U(1) symmetry in eq.(\ref{lattuv}), whose charged operator is $S^{\pm}_j$ (0-dimensional) since $U_z(\theta)S^{\pm}_jU^{\dagger}_z(\theta)=e^{\pm i\theta}S^{\pm}_j$, and the symmetry operator $U_z(\theta)=\rm{exp}(i\theta \sum_j S^{z}_j)$ has 2-dimensional support. 

Similarly, we can consider the Wilson loop operator defined on an arbitrary closed loop, following the construction above.


\section{Constraints from global symmetries in the low energy theory}\label{sectres}
In this section, we consider the global symmetries in the low energy theory which lead to some constraints. We consider a system of size $L\times L$ with $L$ an even number.

\subsection{1-form symmetry and the selection rule}\label{selectionrule}
\begin{figure}
	\centering
	\includegraphics[width=0.48\textwidth]{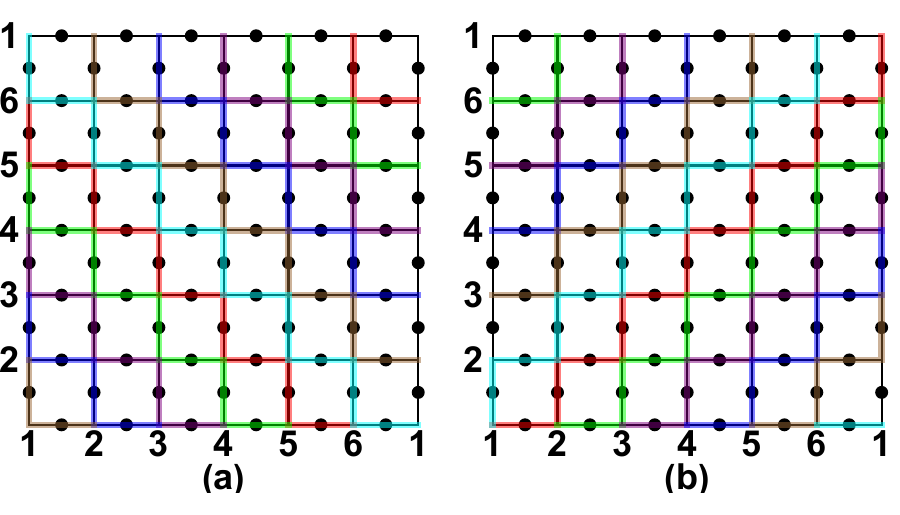}
	\caption{\label{strufact}  Fourier transform of the spin operator $S^{z}_{-\textbf{Q}}$ along two high-symmetry momentum paths. (a) Path along $\textbf{Q}=(q,q)$ and (b) path along $\textbf{Q}=(q,-q)$. In both cases, $S^{z}_{-\textbf{Q}}$ is equivalent to the sum of the Fourier transforms of 1-form conserved quantities $\widetilde{W}^{TLBR}_{M},\widetilde{W}^{BLTR}_{M}$, integrated along the color shaded paths in the square lattice.}
\end{figure}

Information of 1-form symmetry is indeed encoded in the static structure factor $S(\textbf{Q})=\langle S^{z}_{\textbf{Q}}S^{z}_{-\textbf{Q}}\rangle$ and dynamical structure factor $S^{zz}(\textbf{Q},\omega)=\int \frac{dt}{2\pi}e^{i\omega t}\langle S^{z}_{\mathbf{Q}}(t)S^{z}_{-\mathbf{Q}}(0)\rangle$, where $S^{z}_{\mathbf{Q}}=\sum_{\mu=1}^{2}S^{z}_{\mu}(\mathbf{Q})$, and $S^{z}_{\mu}(\mathbf{Q})=\frac{1}{N}\sum_{\textbf{r}}e^{-i\mathbf{Q}\cdot (\mathbf{r}+\mathbf{\rho}_\mu)}S_{\mu}^{z}(\mathbf{r})$. And $\rho_1=\frac{1}{2}e_1$, $\rho_2=\frac{1}{2}e_2$ for the square lattice (Fig.\ref{lattice} (b)). 

i). Consider $\textbf{Q}=(q,q)$ path, 
\begin{equation}
\begin{aligned}\label{qqpathsz}
S^{z}_{-\textbf{Q}}=&\frac{1}{N}\sum_{\textbf{r}}e^{i\textbf{Q}\cdot \textbf{r}}e^{i \frac{q}{2}}[S^{z}_{1}(\textbf{r})+S^{z}_{2}(\textbf{r})]\\
=&\frac{1}{N}\sum_{m,n}e^{iq(m+n)}e^{i \frac{q}{2}}[S^{z}_{1}(m,n)+S^{z}_{2}(m,n)]\\
=&\frac{e^{i\frac{q}{2}}}{N}\sum_{M}e^{iqM}\sum_{m}[S^{z}_{1}(m,M-m)+S^{z}_{2}(m,M-m)]\\
=&\widetilde{W}(q),
\end{aligned}
\end{equation}
where we used the notation $\textbf{r}=m\textbf{e}_1+n\textbf{e}_2\equiv(m,n)$ and $m,n,M$ are integers. $\sum_{m}[S^{z}_{1}(m,M-m)+S^{z}_{2}(m,M-m)]\equiv \widetilde{W}^{TLBR}_{M}$ sums spins along the diagonal loop from top-left to bottom-right as shown in Fig.\ref{strufact} (a). This term can be shown to commute with any plaquette term and with Hamiltonian $H^{\rm{eff}}_{\rm{ice}}$. Consequently, $S^{z}_{-\textbf{Q}}$ for momentum $\textbf{Q}=(q,q)$ can be expressed as a linear combination of these 1-form conserved quantities. Thus, static structure factor measures the Fourier transformation of winding numbers $\widetilde{W}(q), \widetilde{W}(-q)$ along $\textbf{Q}=(q,q)$ path. For the ground states with zero winding numbers, $S(\textbf{Q})$ vanishes along $\textbf{Q}=(q,q)$ path and also along $\textbf{Q}=(q,-q)$ path (see the discussion below). 

As illustrated in Fig.\ref{strufact} (a), we can find that the sum of the spins along two neighboring diagonal loops equals the sum of the ice rule terms under the periodic boundary condition $S^{z}_{1}(m,n-L)=S^{z}_{1}(m,n)$: 
\begin{equation}
\begin{aligned}
&\widetilde{W}^{TLBR}_{M}+\widetilde{W}^{TLBR}_{M-1}\\
=&\sum_{m=1}^{L}[S^{z}_1(m,M-m)+S^{z}_{2}(m,M-m)\\
&+S^{z}_1(m-1,M-m)+S^{z}_2(m,M-m-1)].
\end{aligned}
\end{equation}
As a result, the winding numbers of any two neighboring diagonal loops take opposite values in the ice manifold.

Furthermore, $\widetilde{W}^{TLBR}_{M}$ can be expressed in terms of horizontal loop operators $\widetilde{W}_{\gamma_n}$ and vertical loop operator $\widetilde{W}_{\zeta_m}$:
\begin{equation}
\begin{aligned}\label{hvTLBR}
&\sum_{n=1}^{L}\widetilde{W}_{\gamma_n}+\sum_{m=1}^{L}\widetilde{W}_{\zeta_m}\\
=&\sum_{n=1}^{L}\sum_{\textbf{r}}^{(\gamma_n)}e^{i\textbf{Q}_{0}\cdot \textbf{r}}S^{z}_2(\textbf{r})+\sum_{m=1}^{L}\sum_{\textbf{r}}^{(\zeta_m)}e^{i\textbf{Q}_{0}\cdot \textbf{r}}S^{z}_1(\textbf{r})\\
=&\sum_{M=1}^{L}e^{i\pi M}\widetilde{W}^{TLBR}_{M},
\end{aligned}
\end{equation}
where the loops $\gamma_n$ and $\zeta_m$ pass through $(m,n+\frac{1}{2})$ and $(m+\frac{1}{2},n)$, respectively. In the derivation, we have used the fact that the phase factor $e^{i\textbf{Q}_{0}\cdot \textbf{r}}=e^{i\pi(m+M-m)}=e^{i\pi M}$ for every spin operator in $\widetilde{W}^{TLBR}_{M}$. For the eigenstates in the ice manifold, eq.(\ref{hvTLBR}) leads to the relation among the horizontal, vertical and diagonal winding numbers
\begin{equation}
\widetilde{w}_{\gamma_1}+\widetilde{w}_{\zeta_1}=e^{i\pi M}\widetilde{w}^{TLBR}_{M}.
\end{equation}

For dynamical structure factor, we find 
\begin{equation}
\begin{aligned}\label{dsfeq}
\langle S^{z}_{\mathbf{Q}}(t)S^{z}_{-\mathbf{Q}}(0)\rangle=&\langle e^{iH^{\rm{eff}}_{\rm{ice}}t}S^{z}_{\mathbf{Q}}e^{-iH^{\rm{eff}}_{\rm{ice}}t}S^{z}_{-\mathbf{Q}}\rangle\\
=&\langle e^{iH^{\rm{eff}}_{\rm{ice}}t}\widetilde{W}(q)e^{-iH^{\rm{eff}}_{\rm{ice}}t}\widetilde{W}(-q)\rangle\\
=&\langle \widetilde{W}(q)\widetilde{W}(-q)\rangle,
\end{aligned}
\end{equation}
where we use the result that $\widetilde{W}(q)$ commutes with $H^{\rm{eff}}_{\rm{ice}}$. As a result, the correlator $\langle S^{z}_{\mathbf{Q}}(t)S^{z}_{-\mathbf{Q}}(0)\rangle$ exhibits dependence solely on $q$ along the $\mathbf{Q}=(q,q)$ path, leading to the absence of any finite-frequency response in $S^{zz}(\mathbf{Q},\omega)$. This constitutes a selection rule that constrains the dynamical structure factor, originating from the 1-form symmetry.

ii). Consider $\textbf{Q}=(q,-q)$ path,
\begin{equation}
\begin{aligned}\label{qmqpathsz}
S^{z}_{-\textbf{Q}}=&\frac{1}{N}\sum_{m,n}e^{iq(m-n)}[e^{i \frac{q}{2}}S^{z}_{1}(m,n)+e^{-i \frac{q}{2}}S^{z}_{2}(m,n)]\\
=&\frac{1}{N}\sum_{m,n}e^{iq(m-n)}e^{i \frac{q}{2}}S^{z}_{1}(m,n)\\
&+\frac{1}{N}\sum_{m,n}e^{iq(m+1-n)}e^{-i \frac{q}{2}}S^{z}_{2}(m+1,n)\\
=&\frac{1}{N}\sum_{m,n}e^{iq(m-n)}e^{i \frac{q}{2}}[S^{z}_{1}(m,n)+S^{z}_{2}(m+1,n)]\\
=&\frac{e^{i \frac{q}{2}}}{N}\sum_{M}e^{iqM}\sum_{m}[S^{z}_{1}(m,m-M)+S^{z}_{2}(m+1,m-M)]\\
=&\widetilde{W}'(q),
\end{aligned}
\end{equation}
where translation symmetry and periodic boundary condition have been used in the summation to obtain the second equality. $\sum_{m}[S^{z}_{1}(m,m-M)+S^{z}_{2}(m+1,m-M)]\equiv \widetilde{W}^{BLTR}_{M}$ sums spins along the diagonal loop from bottom-left to top-right as shown in Fig.\ref{strufact} (b). This term can also be shown to commute with any plaquette term and with Hamiltonian $H^{\rm{eff}}_{\rm{ice}}$. The constraints for static and dynamical structure factors are similar to those along the $\textbf{Q}=(q,q)$ path. Our analyses above also remain valid for the models $H^{\rm{LGT}}$, and $H^{\rm{QLM}}$.

Similar to the treatments of the loop operators along the $\textbf{Q}=(q,q)$ path, consider the diagonal loops shown in \ref{strufact} (b), we have 
\begin{equation}
\begin{aligned}
&\widetilde{W}^{BLTR}_{M}+\widetilde{W}^{BLTR}_{M-1}\\
=&\sum_{m=1}^{L}[S^{z}_1(m,m-M)+S^{z}_{2}(m,m-M)\\
&+S^{z}_1(m-1,m-M)+S^{z}_2(m,m-M-1)]
\end{aligned}
\end{equation}
under the periodic boundary condition. Likewise, we find that the winding numbers of any two neighboring diagonal loops take opposite values in the ice manifold. $\widetilde{W}^{TLBR}_{M}$ can also be expressed in terms of horizontal loop operators $\widetilde{W}_{\gamma_n}$ and vertical loop operator $\widetilde{W}_{\zeta_m}$:
\begin{equation}
\begin{aligned}\label{bltreq}
&-\sum_{n=1}^{L}\widetilde{W}_{\gamma_n}+\sum_{m=1}^{L}\widetilde{W}_{\zeta_m}\\
=&-\sum_{n=1}^{L}\sum_{\textbf{r}}^{(\gamma_n)}e^{i\textbf{Q}_{0}\cdot \textbf{r}}S^{z}_2(\textbf{r})+\sum_{m=1}^{L}\sum_{\textbf{r}}^{(\zeta_m)}e^{i\textbf{Q}_{0}\cdot \textbf{r}}S^{z}_1(\textbf{r})\\
=&\sum_{M=1}^{L}e^{-i\pi M}\widetilde{W}^{BLTR}_{M}.
\end{aligned}
\end{equation}

The above results (from eq.(\ref{qqpathsz}) to eq.(\ref{bltreq})) also hold for the following cases: a). system of size $L_1\times L_2$ with $L_1=k L_2$ (where $k$ is an integer) and $L_2$ is even; and b). the spin-S case (see the discussion in appendix \ref{appendix}).


\subsubsection{Application of the constraints from 1-form symmetry to the lattice models}
Since $H^{\rm{eff}}_{\rm{ice}}$ (equivalently,  $H^{\rm{LGT}}$) emerges as the low energy effective theory of the XXZ model in eq.(\ref{lattuv}) and transverse field Ising model \cite{Moessner2001prb,Henry2014prl} near the Ising limit, here we compare above analyses with the previous related numerical studies. As the low energy effective theory is derived via degenerate perturbation theory, ref.\cite{Chernyshev2004prb,Delannoy2005prb} point out that not only the Hamiltonian but also other operators must be transformed to properly describe observables in the low energy theory. In appendix \ref{appendix}, we show that $S^z_i$ remains unchanged from UV to low energy space in the degenerate perturbation theory. Hence, the above constraints on the static structure factor $S(\textbf{Q})$ and the dynamical structure factor $S^{zz}(\textbf{Q},\omega)$ hold within the low energy space of the lattice models whose effective low energy theory is $H^{\rm{eff}}_{\rm{ice}}$.

\emph{XXZ model and transverse field Ising model}. Near the Ising limit on the planar pyrochlore lattice, the ground state of the XXZ model \cite{Ralko2010prl,Xiong2025} (eq.(\ref{lattuv})) and the transverse field Ising model \cite{Moessner2001prb,Henry2014prl} are plaquette valence bond solid (PVBS) phases \cite{Shannon2004prb,Banerjee2013Jstat,Tschirsich2019scipost,Xiaoxue2022,Henry2014prl,Xiong2025,Zoller2014prx}. These phases persist at very low but finite temperatures \cite{Xiaoxue2022,Henry2014prl} and break both lattice translation symmetry and the $C_2$ rotation symmetry with rotation centers located at the midpoints of the links. Moreover, these phases are flat in the height representation \cite{Henley1997,Henley2004,Fradkin2004cantor,Ashvin2004vbs,Moessner2010book} and have zero horizontal and vertical numbers \cite{Tschirsich2019scipost,Xiaoxue2022,Henry2013thesis,Herzog2019prb}. Therefore, the diagonal winding numbers are zero based on eq.(\ref{hvTLBR}) and eq.(\ref{bltreq}). Our constraints rely on well-defined winding numbers, which require the ice rule to hold. At higher temperatures, the ice rule is violated, rendering the winding numbers ill-defined, thus, our constraints no longer apply.

{i). Static structure factor. According to the eq.(\ref{qqpathsz}) and eq.(\ref{qmqpathsz}), the static structure factor $S(\textbf{Q})$ vanishes along $\textbf{Q}=(q,\pm q)$ paths for any state in the ice manifold. This prediction agrees with the pinch-point-like \cite{Gingras2014,Henley2010} structure in $S(\textbf{Q})$ observed numerically in the PVBS phase \cite{Henry2013thesis,Henry2014prl,Xiong2025}. In the transverse field Ising model, the ground state is a canted Neel phase for larger transverse field. And it has been argued that the phase transition between PVBS and canted Neel phases can be captured by the quantum link model. Our prediction is also consistent with the canted Neel phase \cite{Henry2013thesis,Henry2014prl} where Bragg peaks develop. It should be noted that the coordinate system used in those studies differs from ours, a $\pi/4$ rotation is required to transform one into the other. 

{ii). Dynamical structure factor. In ref.\cite{Xiong2025}, the dynamical structure factor $S^{zz}(\textbf{Q},\omega)$ is calculated numerically. In the low energy part ($\omega\ll J_z$), $S^{zz}(\textbf{Q},\omega)$ vanishes along the $(0,0)-(2\pi,2\pi)$ path in our coordinate system, as explained by eq.(\ref{dsfeq}). Along the $(0,0)-(2\pi,0)-(2\pi,2\pi)$ path, it exhibits finite intensity at $\omega\sim 4J_{xy}^2/J_{z}$ and very weak (almost flat) dispersion. This behavior arises because $S^{z}_{-\textbf{Q}}$ does not commute with $H^{\rm{eff}}_{\rm{ice}}$ along this path except at $\textbf{Q}=(0,0)$ or $(2\pi,2\pi)$, and the low energy dynamics are governed by eq.(\ref{ringex}) with energy scale $t=4J_{xy}^2/J_{z}$. Using plaquette operator theory \cite{Langari2015epjb}, ref.\cite{Xiong2025} shows that the dispersion is flat to the lowest order. Higher order corrections could lead to weak dispersion \cite{Brenig2002prb,Auerbach2003prl}.

\emph{Quantum link model}. This model (eq.(\ref{qlm})) has been studied in ref.\cite{Shannon2004prb,Banerjee2013Jstat,Tschirsich2019scipost,Xiaoxue2022}. The phase diagram consists of a Neel phase for $\lambda<-0.35$, a PVBS for $-0.35<\lambda<1$, and a columnar phase for $\lambda>1$. In the height representation \cite{Henley1997,Henley2004,Fradkin2004cantor,Ashvin2004vbs,Moessner2010book,Xiaoxue2022,Banerjee2013Jstat}, both the Neel phase and PVBS phases are flat and have zero horizontal and vertical winding numbers. The columnar phase, by contrast, is tilted with large horizontal and vertical winding numbers and exhibits extensive degeneracy, where many states with different winding numbers become degenerate (the method for generating these states is given in ref.\cite{Shannon2004prb}, also see appendix \ref{appendixb}). According to the above constraints, we expect the static and dynamical structure factors to vanish along the $\textbf{Q}=(q,\pm q)$ paths in the Neel and PVBS phase. For the columnar phase, within a fixed-winding-numbers sector, the dynamical structure factor should also vanishes along these paths for finite $\omega$.


\subsection{LSM type constraints}
Based on the notion of bulk insensitivity \cite{HCPo2017prl,Haruki2018prb,YuanYao2021prl} of the gapped system to the symmetry twisted boundary condition, the LSM theorem has been generalized to discrete symmetries, for example, ref.\cite{HCPo2017prl,YuanYao2022prb}. A key application of bulk insensitivity is that the ground state degeneracy under symmetry twisted boundary conditions implies the degeneracy under periodic boundary conditions in such gapped systems \cite{HCPo2017prl,YuanYao2021prl,YuanYao2022prb}. In fact, it is implicitly assumed that applying a symmetry twisted boundary condition does not introduce boundary modes (that could contribute to degeneracy) in bulk insensitivity. However, it is known that boundary modes may appear in topological phases under the symmetry twisting boundary condition (see examples in ref.\cite{HCPo2017prl,YuanYao2021prl}), therefore, this method does not apply to topological phases.

\subsubsection{Symmetry twisted boundary condition}
Among the symmetries of $H^{\rm{eff}}_{\rm{ice}}$, $H^{\rm{LGT}}$, and $H^{\rm{QLM}}$, we consider the 2-fold rotation symmetry $C_2$ with rotation centers located at the middle point of the links, the Charge conjugation symmetry $C$, and the global U(1) rotation symmetry $U_z(\theta)=\rm{exp}(i\theta \sum_j S^{z}_j)$ from local U(1) symmetry. It is sufficient to consider the $Z_2$ rotation $U_z(\pi)=\rm{exp}(i\pi \sum_j S^{z}_j)$, which is a special case of $U_z(\theta)$. 

Following ref.\cite{HCPo2017prl,YuanYao2022prb}, we choose a path $\gamma$ (see Fig.\ref{twist}(a)) and locally modify the Hamiltonians along $\gamma$. Each plaquette term $H_\textbf{r}=O^{L}_{\textbf{r}}O^{R}_{\textbf{r}}$ intersecting $\gamma$ is twisted by the $Z_2$ rotation symmetry $U_z(\pi)$: $H^{\rm{twisted}}_\textbf{r}=O^{L}_{\textbf{r}}U_z(\pi) O^{R}_{\textbf{r}} U_z^{-1}(\pi)$. The operators $O^{L}_{\textbf{r}}$ and $O^{R}_{\textbf{r}}$ are localized to the left and right of $\gamma$, respectively—a well-defined notion due to the finite-range interaction. And only the right operators are twisted. Given that $H^{\rm{eff}}_{\rm{ice}}$, $H^{\rm{LGT}}$, and $H^{\rm{QLM}}$ possess identical global symmetries, we restrict our analysis to $H^{\rm{eff}}_{\rm{ice}}$ without loss of generality. The ice rule terms are unchanged under the twist (this is different from the toric code model discussed in the supplemental materials of ref.\cite{HCPo2017prl}). Consider a plaquette term $H_\textbf{r}=-t(S^{+}_iS^{-}_jS^{+}_kS^{-}_l+\text{h.c.})$ at site $\textbf{r}$ (see Fig.\ref{twist}(a)). Relative to the path $\gamma$, site $i$ lies to the left and sites $j,k,l$ to the right. Then, $H^{\rm{twisted}}_\textbf{r}=-H_\textbf{r}$. The full twisted Hamiltonian $H^{\rm{twisted}}_\gamma$ is shown in Fig.\ref{twist}(a), with twisted plaquettes shaded red. 

\begin{figure}
	\centering
	\includegraphics[width=0.45\textwidth]{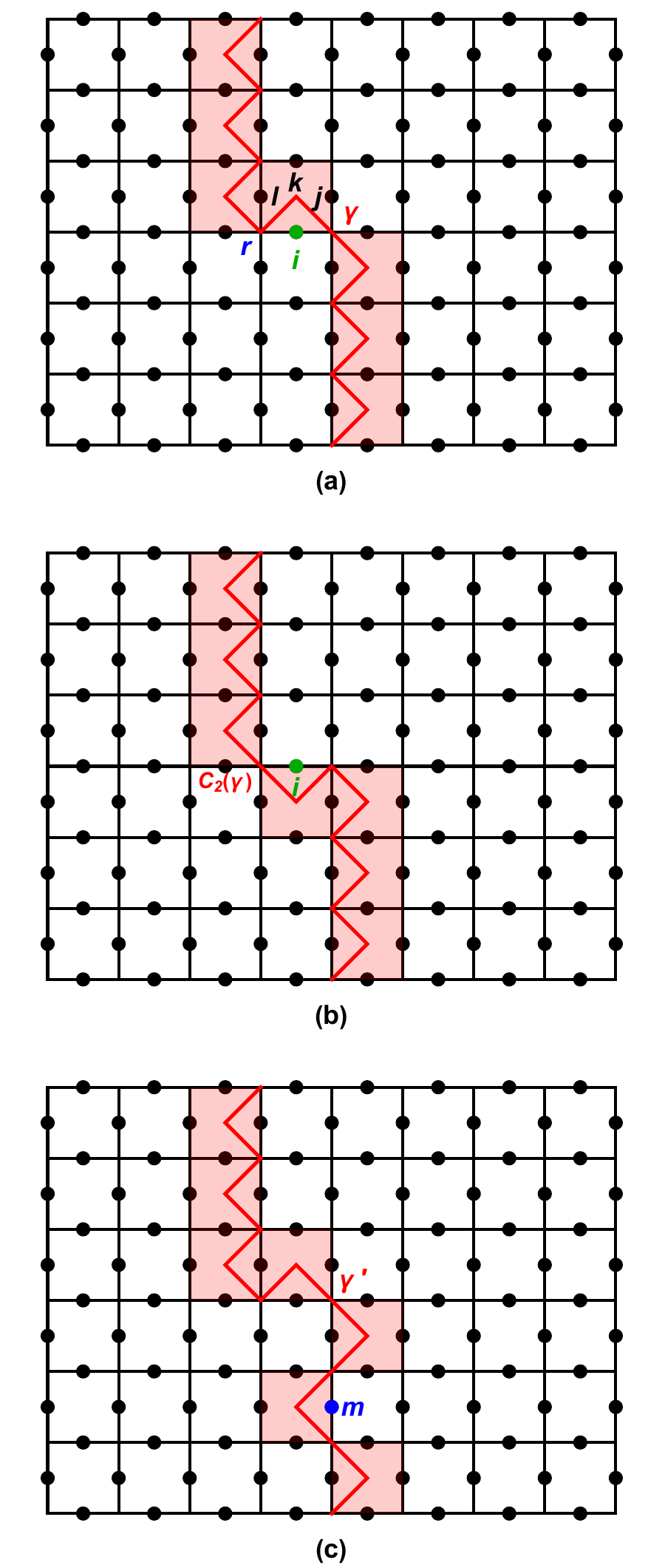}
	\caption{\label{twist}  (a) Twisted Hamiltonian $H^{\rm{twisted}}_\gamma$ with twisted path $\gamma$. The twisted plaquette terms are shaded in red. (b) $H^{\rm{twisted}}_\gamma$ transformed by a 2-fold rotation symmetry about site $i$: $C_2 H^{\rm{twisted}}_\gamma C_2 ^{-1}=H^{\rm{twisted}}_{C_2(\gamma)}$. (c) The twisted path $\gamma$ can be deformed by a unitary transformation: $\rm{exp}(i\pi S^{z}_m)H^{\rm{twisted}}_\gamma\rm{exp}(-i\pi S^{z}_m)=H^{\rm{twisted}}_{\gamma'}$.}
\end{figure}

 It is straightforward to verify that the twisted Hamiltonian respects the charge conjugation symmetry $C$ (eq.(\ref{charcon})). However, it explicitly breaks the original 2-fold rotation symmetry (with center at site $i$), as the twisted path becomes $C_2(\gamma)$, see Fig.\ref{twist} (b). Nevertheless, the twisted Hamiltonian $H^{\rm{twisted}}$ exhibits a modified 2-fold rotation symmetry $C_2 '=\rm{exp}(i\pi S^{z}_i)\,C_2$, satisfying 
\begin{equation}
C_2 'H^{\rm{twisted}}_\gamma (C_2 ')^{-1}=H^{\rm{twisted}}_\gamma.
\end{equation}
 Furthermore, one finds that
\begin{equation}
CC_{2}'C^{-1}=(-1)^{2S}C_2 ',
\end{equation}
where the minus sign arises from $\rm{exp}(i\pi S^{z}_i)\rm{exp}(i\pi S^{x}_i)=(-1)^{2S}\rm{exp}(i\pi S^{x}_i)\rm{exp}(i\pi S^{z}_i)$ for spin-S. It follows from the above equation that every eigenstate of $H^{\rm{twisted}}_\gamma$ is at least twofold degenerate, as $S=1/2$ is considered in this paper}. Consequently, the bulk insensitivity dictates that any gapped phase described by the untwisted model (effective model Eq. (\ref{ringice}), U(1) lattice gauge theory Eq. (\ref{lgt}), or the spin-1/2 U(1) quantum link model Eq. (\ref{qlm}) ) must exhibit a ground state degeneracy. In other words, none of these Hamiltonians can host a symmetric, unique gapped ground state. Since local gauge symmetries can never be spontaneously broken \cite{Elitzur1975}, under the assumption of bulk insensitivity, any gapped phase must necessarily break at least one of the remaining symmetries implicated by above LSM type constraint, namely, charge conjugation or 2-fold rotation symmetry. Alternatively, a symmetric gapped phase could be topologically ordered, but such a scenario may not be captured by the bulk insensitivity argument.

Above LSM type constraint is consistent with the phase diagram of U(1) quantum link model \cite{Shannon2004prb,Banerjee2013Jstat,Tschirsich2019scipost,Xiaoxue2022}, where the phases are either gapped and symmetry broken or gapless. For example, the Neel phase breaks charge conjugation and translation symmetry, whereas the PVBS phase restores charge conjugation symmetry but breaks 2-fold rotation symmetry and translation symmetry. 

The twisted path $\gamma$ above is not unique, it can be deformed by unitary transformation, for example, $\rm{exp}(i\pi S^{z}_m)H^{\rm{twisted}}_\gamma\rm{exp}(-i\pi S^{z}_m)=H^{\rm{twisted}}_{\gamma'}$, see Fig.\ref{twist} (c). 

\subsubsection{1-form symmetry defect}
The framework of generalized global symmetries \cite{Gaiotto2015} introduces the idea of symmetry defects as local operators that modify the system. For a $p$-form symmetry in $(d+1)$ dimensional spacetime, the defect is a $(d-p)$-dimensional spacetime object extended in time. In the Hamiltonian picture, this becomes a local modification acting on a $(d-p-1)$-dimensional spatial region. Therefore, in a two dimensional lattice system, the defect is one-dimensional for a 0-form symmetry. The symmetry twisted boundary condition described above is equivalent to the symmetry defect approach \cite{Gaiotto2015,MengCheng2022,Choi2025,Sahand2024}, where a 0-form $Z_2$ symmetry defect is inserted along a noncontractible cycle \cite{Choi2025}. 

 We now consider the defect Hamiltonian for the U(1) 1-form symmetry, whose corresponding symmetry defect is 0-dimensional. Following ref.\cite{Choi2025}, we first define an interval operator on an open interval $\omega$ (see the red path in Fig.\ref{twist1f}) by cutting open the symmetry operator. Specifically, we choose the path shown in Fig.\ref{strufact} (b) to define a 1-form U(1) symmetry operator, then the interval operator is given by
\begin{equation}
\eta(\omega)=\rm{exp}\left(i\alpha \sum_{m\in \omega}[S^{z}_{1}(m,m-M)+S^{z}_{2}(m+1,m-M)]\right),
\end{equation}
where $\alpha$ is a U(1) parameter. Then, we conjugate the original Hamiltonian (again, we consider $H^{\rm{eff}}_{\rm{ice}}$) by the interval operator $\eta(\omega)$: $\eta(\omega) H^{\mathrm{eff}}_{\mathrm{ice}} \eta^{-1}(\omega)$. This transformation creates a pair of 1-form symmetry defects located at the endpoints of the open interval $\omega$. The defect Hamiltonian $H^{\mathrm{defect}}$ for a single defect located at site $\textbf{r}+\textbf{e}_2$ [see Fig. \ref{twist1f}(b)] is then obtained by modifying only one plaquette term of the original Hamiltonian. This modified term, corresponding to the plaquette shaded red in Fig.\ref{twist1f} (b), is given by 
\begin{equation}
S^{+}_{1}(\textbf{r})S^{-}_{2}(\textbf{r}+\textbf{e}_1)e^{i\alpha}S^{+}_{1}(\textbf{r}+\textbf{e}_2)S^{-}_{2}(\textbf{r})+h.c.
\end{equation}
where we have used $e^{i\alpha S^{z}_m}S^{\pm}_m e^{-i\alpha S^{z}_m}=e^{\pm i\alpha}S^{\pm}_m$.

The defect Hamiltonian has charge conjugation symmetry when $\alpha=\pi$, as
\begin{equation}
\begin{aligned}
C:& S^{+}_{1}(\textbf{r})S^{-}_{2}(\textbf{r}+\textbf{e}_1)e^{i\alpha}S^{+}_{1}(\textbf{r}+\textbf{e}_2)S^{-}_{2}(\textbf{r})\\
&+S^{-}_{1}(\textbf{r})S^{+}_{2}(\textbf{r}+\textbf{e}_1)e^{-i\alpha}S^{-}_{1}(\textbf{r}+\textbf{e}_2)S^{+}_{2}(\textbf{r})\\
\to&S^{-}_{1}(\textbf{r})S^{+}_{2}(\textbf{r}+\textbf{e}_1)e^{i\alpha}S^{-}_{1}(\textbf{r}+\textbf{e}_2)S^{+}_{2}(\textbf{r})\\
&+S^{+}_{1}(\textbf{r})S^{-}_{2}(\textbf{r}+\textbf{e}_1)e^{-i\alpha}S^{+}_{1}(\textbf{r}+\textbf{e}_2)S^{-}_{2}(\textbf{r}).
\end{aligned}
\end{equation}
Additionally, the defect Hamiltonian possesses a modified 2-fold rotation symmetry $C_{2}'=\rm{exp}(i\pi S^{z}_p) C_2$ where the rotation center is located at site p (see Fig.\ref{twist1f} (b)). One can also verify that $CC_{2}'C^{-1}=(-)^{2S}C_2 '$, which implies the ground state degeneracy for half-integer spin. 

\begin{figure}
	\centering
	\includegraphics[width=0.4\textwidth]{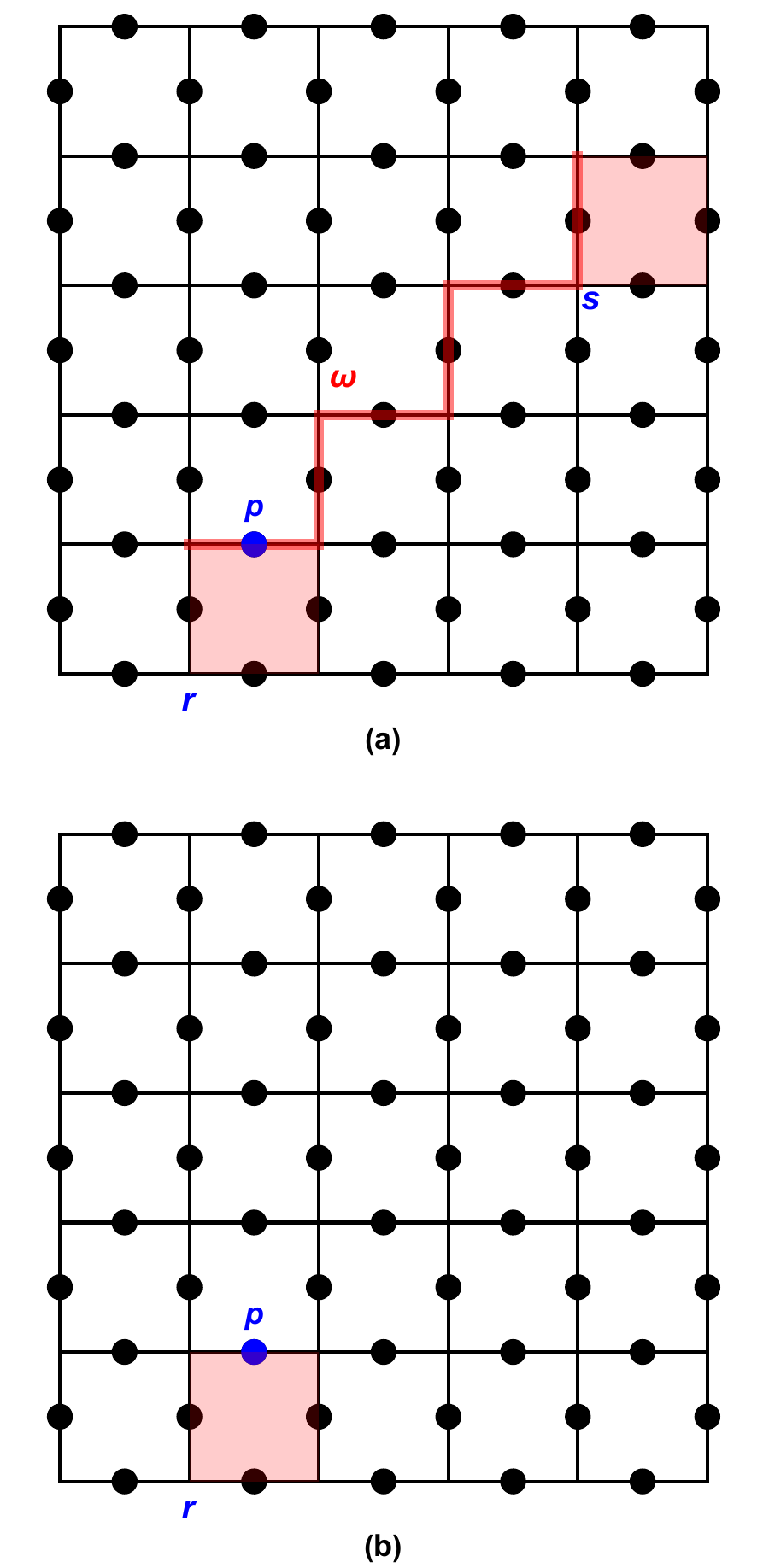}
	\caption{\label{twist1f} 1-form symmetry defect. (a) Interval operator defined on the open red path $\omega$. (b) Defect Hamiltonian considered in the main text. The plaquettes shaded in red indicate the plaquette term modified by the symmetry defect.}
\end{figure}

Usually, a U(1) 1-form symmetry cannot be spontaneously broken in (2+1)-dimensional spacetime \cite{Lake2018}, in accordance with a generalized version of the Mermin-Wagner theorem. Consequently, any gapped phase in our models must break either charge-conjugation or two-fold rotation symmetry or could be topological ordered phase. This is consistent with the results obtained above.

\subsubsection{Discussion of the LSM type constraints}
Above LSM type constraints are derived from the low energy effective theory, in this part, we discuss the LSM type constraints in the XXZ model (eq.(\ref{lattuv})). 

Besides the original LSM theorem \cite{LSM1961}, whose proof is based on a variational argument and a unitary twist operator, modern LSM type constraints can be classified into two classes \cite{Tasaki2022review,Sahand2024,Zou2026}. The first class relies on the nontrivial projective representation of the symmetry \cite{YuanYao2021prl,HCPo2017prl,YuanYao2022prb,Haruki2015pnas,XChen2011prb} and applies to the systems with internal symmetries including discrete ones. The second class applies to the systems with U(1) symmetry and is based on the fractional filling factor \cite{Oshikawa2000prl,Oshikawa1997prl}. Since the U(1) group admits no nontrivial projective representations, the second class works for the systems whose internal symmetry is solely U(1), whereas the first class typically provides no constraint in such systems.

In addition to the spatial symmetry of the planar pyrochlore lattice, pure XXZ model (eq.(\ref{lattuv}) with $h=0$) possesses $U(1)\rtimes Z_{2}$ symmetries and time reversal symmetry $Z_2^T$. In this case, both the projective representation method and filling factor method apply. When $h\neq 0$, only the U(1) symmetry remains, and consequently only the filling factor method applies.

i).LSM constraints for XXZ model based on the projective representation method. 
\begin{itemize}
\item Since the spins are located at the $C_2$ rotation centers of the lattice, and there is nontrivial projective representation of $Z_2\times Z_2$ symmetries (a subgroup of $U(1)\rtimes Z_{2}$) for half-integer spin, a symmetric, unique gapped ground state is forbidden by lattice homotopy \cite{HCPo2017prl} or symmetry twisted boundary method \cite{YuanYao2022prb}. 
\item Recently, ref.\cite{Tada2026prb} shows that a symmetric, unique gapped ground state is forbidden under the periodic boundary condition in the half-integer spin XXZ model with $U(1)\rtimes Z_{2}$ and point group $C_{2v}$ symmetries by introducing a U(1) flux to twist the point group symmetries. 
\item Ref.\cite{Tada2026prb} also provides an LSM constraint in the same model by considering $U(1)\times Z_2^T$ and $C_2$ symmetries.
\end{itemize} These constraints allow for phases with broken spin symmetry (which, due to the continuous nature of the spin symmetry, may result in a gapless phase) or broken spatial symmetry ($C_2$, or $C_{2v}$). For gapped phases near the Ising limit, these constraints agree with the LSM constraints from the low energy theory.

ii).LSM constraints for eq.(\ref{lattuv}) based on the filling factor method. As there are two spins in the unit cell, ref.\cite{Furuya2019prb} shows that the U(1) flux insertion argument \cite{Oshikawa2000prl} provides no LSM constraint for the model with U(1) symmetry on the planar pyrochlore lattice under the periodic boundary condition or tilted boundary condition \cite{YuanYao2020prx} when the ground state has zero total magnetization, as in the case for the quantum ice. By constructing a tilted Klein bottle boundary condition and using the U(1) symmetry together with a translation symmetry, ref.\cite{Furuya2019prb} derives an LSM constraint. However, this boundary condition is valid only for systems of size $L\times L$, as it relies on the $C_4$ rotation and reflection symmetries.

For the quantum ice regime we considered in this paper, i.e. XXZ model with a Zeeman field ($-\frac{1}{2}<\frac{h}{2J_z}<\frac{1}{2}$) near the Ising limit, a symmetric, unique gapped ground state is prohibited when the system is compatible with the tilted Klein bottle boundary condition. On the other hand, we see that the symmetric, unique gapped ground state can be prohibited even in the system with lower spatial symmetries from the low energy effective theory, where only $C_2$ rotation is used. This implies that an LSM constraint persists even when the spatial symmetry of the model is reduced from $C_4$ to $C_2$.

For the quantum link model, we note that there exists another LSM constraint based on the filling factor method, which generalizes the original 0-form U(1) symmetry in the flux insertion argument \cite{Oshikawa2000prl} to a 1-form U(1) symmetry \cite{Kobayashi2019prb}. However, the filling of the 1-form charge cannot be directly read off from the model, and may take different values in different phases. Therefore, once the phase diagram is determined, this LSM constraint can be used to check the consistency between the symmetry-breaking patterns in each phases and the corresponding filling. In contrast, the symmetry twisted boundary condition and symmetry defect method used in the previous section do not require such a posteriori information.


\section{Conclusions}
In this work, we study the low energy effective theory $H^{\rm{eff}}_{\rm{ice}}$ of quantum ice on the planar pyrochlore lattice and also the closely related U(1) quantum link model from the perspective of their global symmetries. Our main findings are as follows.

i). Along two high-symmetry momentum paths $\textbf{Q}=(q,\pm q)$, the Fourier transform of the spin operator $S^{z}_{-\textbf{Q}}$ can be expressed in terms of 1-form conserved quantities. Consequently, the static structure factor measures the winding number along these two paths. Furthermore, this leads to a complete absence of finite-frequency response in the dynamical structure factor $S^{zz}(\mathbf{Q},\omega)$ for $H^{\rm{eff}}_{\rm{ice}}$. 

ii). Utilizing the symmetry twisted boundary condition and the notion of bulk insensitivity, we derive two LSM type constraints that forbids the existence of a unique gapped ground state of these models. These constraints are consistent with previous numerical studies \cite{Shannon2004prb,Banerjee2013Jstat,Tschirsich2019scipost,Xiaoxue2022}.

We expect that the constraints we have uncovered can be applied to more realistic experimental settings in quantum simulators, as the low energy effective models obtained by perturbing the Ising model on the planar pyrochlore lattice share the same symmetries.

 
\begin{acknowledgements}
We thank Yuan Yao, Linhao Li and Yunqin Zheng for helpful discussions. This work is supported by the National Science Foundation of China: No. 12404169, No. 12147172, the International Postdoctoral Exchange Fellowship Program 2022 by the Office of China Postdoctoral Council: No.PC2022072 and the start-up funding of Chongqing Normal University (Grant No. 24XLB010).
\end{acknowledgements}

\appendix
\section{Degenerate perturbation theory}\label{appendix}
Here, we briefly introduce the degenerate perturbation theory used in ref.\cite{Kato1949,Klein1974,Takahashi1977}. Consider the Hamiltonian
\begin{equation}
H=H_0+\lambda V
\end{equation}
where $H_0$ is assumed to be solvable and has an $m$-fold degenerate ground states manifold spanning the space $\mathcal{H}_0$. The ground state energy of $H_0$ is denoted by $E_0$. Here we consider the systems in which there is a large gap $\Delta$ between the ground state and the first excited state. The quantum ice studied in the main text clearly belongs to this class. $\lambda V$ is the perturbation term, with $\lambda$ being a dimensionless parameter. When the perturbation is turned on, the $m$-fold degeneracy of the ground states of $H_0$ may be lifted, causing their energies to split. Let $\delta$ denote the maximum energy difference among these $m$ perturbed states, and let $\mathcal{H}$ be the space spanned by these $m$ states. A Large gap means $\Delta\gg\delta$. Finally, we define $P_0$ and $P$ as the projection operators onto the space $\mathcal{H}_0$ and $\mathcal{H}$, respectively.

The projection operator $P$ can be written as
\begin{equation}
P=\frac{1}{2\pi i}\oint_C \frac{1}{z-H}dz, 
\end{equation}
where the contour $C$ is chosen to enclose only the eigenenergies of those $m$ perturbed states. To derive the perturbation series expansion, we iteratively use the identity and obtain
\begin{equation}
\begin{aligned}
\frac{1}{z-H}=&\frac{1}{z-H_0}(z-H+\lambda V)\frac{1}{z-H}\\
=&\frac{1}{z-H_0}(1+\lambda V\frac{1}{z-H})\\
=&\frac{1}{z-H_0}\sum_{m=0}^{\infty}\lambda^m(V\frac{1}{z-H_0})^m.
\end{aligned}
\end{equation}
We can decompose $1/(z-H_0)$ using the projection operators
\begin{equation}
\frac{1}{z-H_0}=\frac{1}{z-E_0}P_0+(1-P_0)\frac{1}{z-H_0},
\end{equation}
and define $S^0(z)=\frac{1}{z-E_0}P_0, S(z)=(1-P_0)\frac{1}{z-H_0}$. Substituting into the expansion of $P$, we have
\begin{equation}
P=\frac{1}{2\pi i}\sum_{n=0}^\infty \lambda^n \sum_{[k_i]_n}\oint_C S^{k_1}(z)VS^{k_2}(z)V\cdots VS^{k_{n+1}}(z),
\end{equation}
where $k_i$ takes 0 or 1, and $[k_i]_n$ is short for $k_1,k_2,k_3,\cdots, k_{n+1}$. The contour integral can be evaluated by the residue theorem, finally, the projection operator is given by
\begin{equation}
P=P_0-\sum_{n=1}^\infty \lambda^n \sum_{\{k_i\}_n}S^{k_1}VS^{k_2}V\cdots VS^{k_{n+1}},
\end{equation}
where the sum is over all nonnegative integers $k_i$ satisfying the condition: $k_1+k_2+\cdots +k_{n+1}=n$. Here, $S^0=-P_0$, and for $k\geq 1$, $S^k=(1-P_0)\frac{1}{(E_0-H_0)^k}$.

Since a state $\psi_0$ in the space $\mathcal{H}_0$ is easy to construct, we need a transformation that maps it to the corresponding perturbed state $\psi$ in $\mathcal{H}$. A naive choice is $\tilde{U}=PP_0$. Using the property of projection operator $P^2=P$, we have $(\tilde{U})^{\dagger}\tilde{U}=P_0PP_0$. To preserve the inner product under the transformation, we can consider a "normalized" transformation $U=PP_0(P_0PP_0)^{-1/2}$. Then, we have $U^{\dagger}U=P_0$, and $(\psi,\phi)=(U\psi_0, U\phi_0)=(\psi_0,U^\dagger U\phi_0)=(\psi_0,\phi_0)$ for any $\psi_0,\phi_0\in \mathcal{H}_0$. The definition of $(P_0PP_0)^{-1/2}$ is defined by
\begin{equation}
(P_0PP_0)^{-1/2}=P_0+\sum_{n=1}^\infty \frac{(2n-1)!!}{(2n)!!}[P_0(P_0-P)P_0]^n.
\end{equation}

Using the transformation defined above, we can construct an effective Hamiltonian on space $\mathcal{H}_0$
\begin{equation}
\begin{aligned}
H^{\rm{eff}}=&U^{\dagger}HU\\
=&E_0P_0+\lambda (P_0PP_0)^{-1/2}VPP_0(P_0PP_0)^{-1/2},
\end{aligned}
\end{equation}
where we have used $HP=PH$. For the eigenvalue problem in space $\mathcal{H}$: $H\psi=E\psi$, the corresponding equation in space $\mathcal{H}_0$ is $H^{\rm{eff}}\psi_0=E\psi_0$. The effective Hamiltonian derived by this method up to 4th order is reported in ref.\cite{Hartmann2002epjb}:
\begin{equation}\label{dptsec}
H^{\rm{eff}}=E_0P_0+\lambda P_0 V P_0+\lambda^2 P_0 V S V P_0+\cdots. 
\end{equation}
Another common approach for deriving effective theory is the canonical transformation \cite{MacDonald1988prb}. For the single band half-filled Hubbard model \cite{Chernyshev2004prb}, we have verified that the effective Hamiltonian obtained from these two methods agree at least up to 4th order.

\emph{XXZ model.} For the spin-1/2 XXZ model under a Zeeman field (eq.(\ref{lattuv})) with $|J_{xy}|\ll J_z$ and $-\frac{1}{2}<\frac{h}{2J_z}<\frac{1}{2}$, $P,P_0$ are the projection operators onto the space satisfying the ice rule. $V=-J_{xy}\sum_{\langle i,j\rangle}(S^{+}_{i}S^{-}_{j}+S^{-}_{i}S^{+}_{j})$. Using eq.(\ref{dptsec}), the lowest nonzero correction arises at second order and gives eq.(\ref{ringex}). Higher order corrections are reported in ref.\cite{Henry2014prl}.

Recently, the spin ice problem has been generalized to spin-S cases \cite{Damle2025,Haruki2026ice1,Haruki2026duality,Haruki2026ice2}. Inspired by these studies, we consider the spin-S quantum ice on the planar pyrochlore lattice. Near the Ising limit, the ground state manifold remains constrained by the ice rule: $\sum_{i\in\boxtimes}S_{i}^{z}=0$. And the lowest order effective theory from eq.(\ref{dptsec}) takes the same form as eq.(\ref{ringex}). In this case, the mapping from the low energy spin model to lattice gauge theory should be modified. Nevertheless, using the low energy spin model directly, one can find that the 1-form conserved quantities and the constraints on the  static and dynamical structure factor still apply, as only the algebra of spin angular momentum was used to derive these constraints.

\emph{Operators in the low energy space under the degenerate perturbation theory.} It is pointed out in ref.\cite{Chernyshev2004prb,Delannoy2005prb} that in the low energy effective theory, all operators should also be transformed along with the Hamiltonian: $O^{\rm{eff}}=U^{\dagger}OU$. They found corrections from perturbation theory for an operator $O$ in low energy space. Specifically, the staggered magnetization in the Hubbard model is not the same as the magnetic moment operator in the low energy spin model. We have verified that the corrections derived using the above method are identical to those in ref.\cite{Chernyshev2004prb,Delannoy2005prb}.

For the operator $S^z_i $ considered in Section.\ref{selectionrule}, the effective operator in the low energy space up to second order is given by
\begin{equation}
\begin{aligned}
U^{\dagger}S^z_iU=&P_0S^z_i P_0+\lambda^2(P_0VS S^z_i SVP_0-\frac{1}{2}P_0 S^z_i P_0 VS^2 VP_0\\
&-\frac{1}{2}P_0VS^2 VP_0S^z_i P_0+\cdots)
\end{aligned}
\end{equation}
where $V=-J_{xy}\sum_{\langle i,j\rangle}(S^{+}_{i}S^{-}_{j}+S^{-}_{i}S^{+}_{j})$ for the spin-1/2 XXZ model under a Zeeman field. A careful calculation shows that the second order correction terms cancel out exactly. 

Actually, the correction to $S^z_i $ in the low energy space vanishes can be understood on a more general grounds.  
Consider the action of $S^z_i$ on the ice rule state: it does not create excitation that violate the ice rule. Consequently, the matrix elements of $S^z_i$ between an ice rule state and a non-ice rule state are zero. Equivalently, $(1-P)S^z_i P=0$ and $PS^z_i (1-P)=0$. This implies $PS^z_i=S^z_i P$. Similarly, $P_0S^z_i=S^z_i P_0$. It then follows that the low energy effective operator for $S^z_i$: $U^{\dagger}S^z_iU=S^z_i$, i.e. $S^z_i $ remains unchanged from UV to IR under this perturbation theory.

\section{Columnar phase of U(1) quantum link model}\label{appendixb}
In this part, we discuss the phase of the quantum link model (eq.(\ref{qlm})) for $\lambda >1$. This phase is referred to as the columnar phase in ref.\cite{Tschirsich2019scipost}, the quasi-collinear phase in ref.\cite{Shannon2004prb} and the staggered phase in ref.\cite{Xiaoxue2022}.

To visualize the configurations in this phase \cite{Shannon2004prb,Tschirsich2019scipost}, it is convenient to represent spin up (down) using arrows: on horizontal links, spin up (down) is represented by a rightward (leftward) arrow, and on vertical links, by an upward (downward) arrow. The arrows actually represent the $E_{\mu}(\textbf{r})$ fields. Since the ring-exchange terms and $f_{\textbf{r}}$ are related via a unitary transformation given in the main text, transforming $S^{z}_{\mu}(\textbf{r})$ yields $e^{i\textbf{Q}_{0}\cdot \textbf{r}}S^{z}_{\mu}(\textbf{r})$. In this representation, the Gauss law implies "two-in-two-out" rule: at every vertex of the square lattice, two arrows point inward and two arrows point outward. The operator $f_{\textbf{r}}=S^{+}_{1}(\textbf{r})S^{+}_{2}(\textbf{r}+\textbf{e}_1)S^{-}_{1}(\textbf{r}+\textbf{e}_2)S^{-}_{2}(\textbf{r})+h.c.$ acts as follows. It flips flippable plaquettes --- converting a clockwise configuration ($|\circlearrowright\rangle$) on a plaquette to a counter-clockwise one ($|\circlearrowleft\rangle$) and vice versa, see Fig.\ref{apcou} (a), while projecting out all other configurations. And, $f_{\textbf{r}}^2$ counts the number of flippable plaquettes. Consequently, eq.(\ref{qlm}) can be rewritten in the Rokhsar-Kivelson form Hamiltonian
\begin{equation}
\begin{aligned}\label{RKQLM}
H^{\rm{QLM}}=&\sum_{\square}\bigg[-\biggl(|\circlearrowleft\rangle\langle\circlearrowright|+|\circlearrowright\rangle\langle\circlearrowleft|\biggr)+\lambda\biggl(|\circlearrowleft\rangle\langle\circlearrowleft|+|\circlearrowright\rangle\langle\circlearrowright|\biggr)\bigg]\\
=&\sum_{\square}\bigg[\biggl(|\circlearrowleft\rangle-|\circlearrowright\rangle\biggr)\biggl(\langle\circlearrowleft|-\langle\circlearrowright|\biggr)\\
&+(\lambda-1)\biggl(|\circlearrowleft\rangle\langle\circlearrowleft|+|\circlearrowright\rangle\langle\circlearrowright|\biggr)\bigg]. 
\end{aligned}
\end{equation}

\begin{figure}
	\centering
	\includegraphics[width=0.5\textwidth]{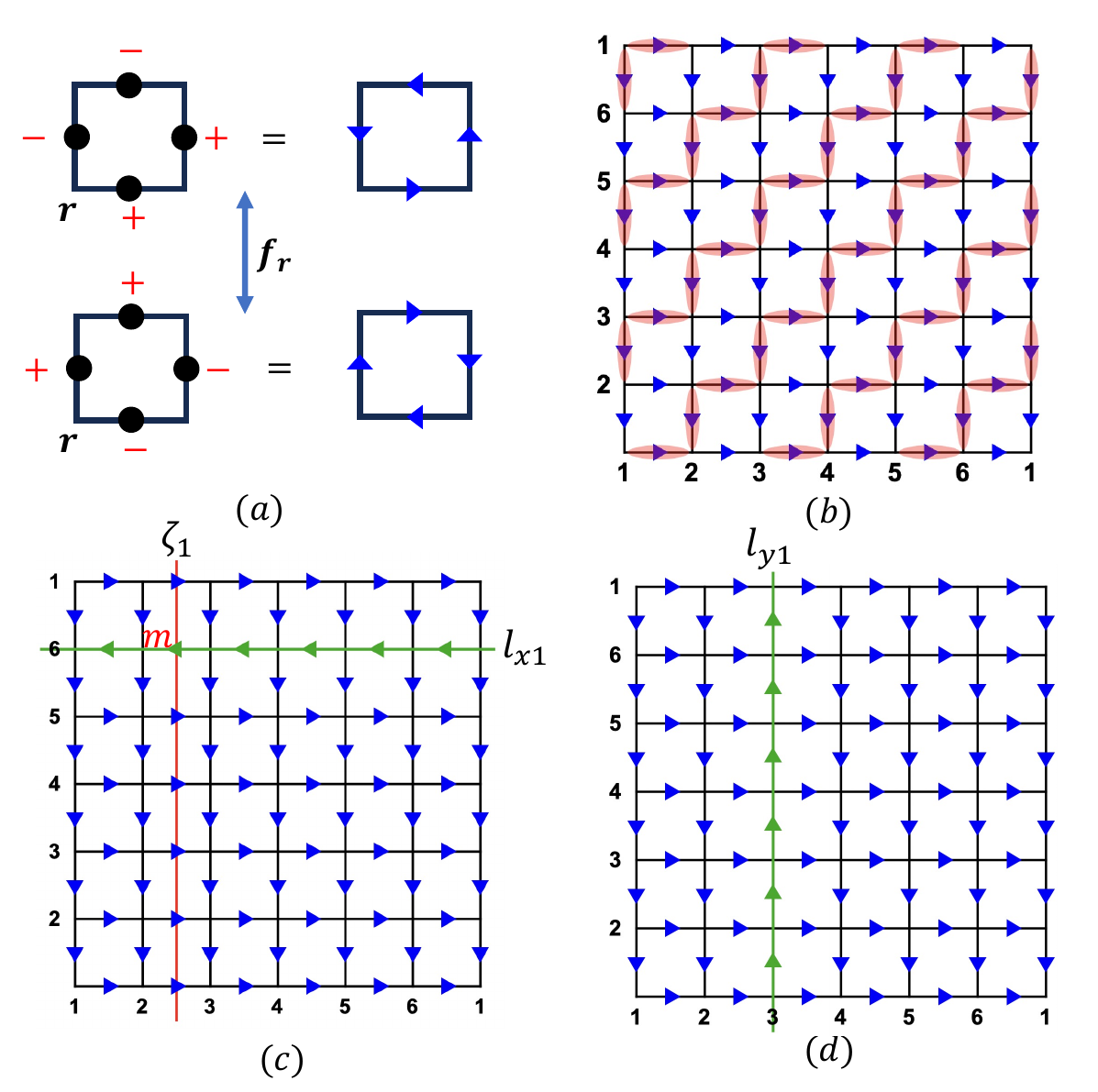}
	\caption{\label{apcou} (a) Arrow representation of spin up (down) in the quantum link model: on horizontal links, spin up (down) is representing by rightward (leftward) arrow; on vertical links, spin up (down) is representing by upward (downward) arrow. The operator $f_{\textbf{r}}$ converts a flippable configuration (top right) to another (bottom right). (b) Configuration of a reference state $|(-L_x/2, L_y/2)\rangle$. Red ovals (empty link) denote spin up (down) following the quantum spin ice model convention. The unitary transformation connecting these two models is given in the main text. (c),(d) The operators $C_{l_k}$ are defined on the straight lines (noncontractible loops), such as $l_{x1},l_{y1}$. }
\end{figure}

For the parameter range $\lambda>1$ considered here, this Hamiltonian becomes positive semidefinite \cite{Moessner2010book} and thus has only non-negative eigenvalues. Accordingly, it is straightforward to verify that any configuration containing no flippable plaquettes \cite{Shannon2004prb,Tschirsich2019scipost} has zero eigenvalues and hence constitutes a ground state of eq.(\ref{RKQLM}) for $\lambda>1$. Four reference states without any flippable plaquettes can be easily constructed. They exhibit "polarized" configurations, in which all horizontal arrows point either rightward or leftward, and all vertical arrows point either downward or upward. One such configuration is shown in Fig.\ref{apcou} (b), this state breaks the charge conjugation symmetry. It should be noticed that this state preserves translation symmetries along $\textbf{e}_1,\textbf{e}_2$ in the quantum link model, while it breaks them in the quantum ice model.

Let the winding numbers of a ground state be denoted by $(\widetilde{\omega}_{\gamma},\widetilde{\omega}_{\zeta})$, where $\widetilde{\omega}_{\gamma},\widetilde{\omega}_{\zeta}$ are the horizontal and vertical winding numbers defined in eq.(\ref{horverwinding}), respectively. Then, for a system of size $L_x\times L_y$, the four "polarized" reference states have winding numbers $(\pm\frac{L_x}{2},\pm \frac{L_y}{2})$ and can be denoted by $|(\pm\frac{L_x}{2},\pm \frac{L_y}{2})\rangle$.

Ref.\cite{Shannon2004prb} provides a method for generating states without any flippable plaquettes from the reference states. We formalize this method as follows. Define unitary transformation along straight line $l_x$ and $l_y$ (they are noncontractible loops, see $l_{x1},l_{y1}$ in Fig.\ref{apcou} (c),(d)) parallel to $\textbf{e}_1$ and $\textbf{e}_2$, respectively,
\begin{equation}
C_{l_k}=\prod_{i\in l_k}e^{i\pi S^{x}_i},
\end{equation}
$k=x,y$. Transformations defined on different lines commute with each other. However, they do not commute with $H^{\rm{QLM}}$. Acting on a reference state, $C_{l_k}$ flips the direction of every arrow on $l_k$, maintains the "two-in-two-out" rule, and avoids generating flippable plaquettes. The resulting states $C_{l_k}|(\pm\frac{L_x}{2},\pm \frac{L_y}{2})\rangle$ still contain no flippable plaquettes (see Fig.\ref{apcou} (c),(d)), thus they are the ground states. Let $|\psi\rangle$ be a reference state with vertical winding number $\widetilde{\omega}_{\zeta}$. Suppose the vertical line $\zeta$ intersects the horizontal line $l_{x1}$ at the link $(\textbf{m},\textbf{m}+\textbf{e}_1)$ (see Fig.\ref{apcou} (c)). Then we can find
\begin{equation}
\begin{aligned}
\widetilde{W}_{\zeta}C_{l_{x1}}|\psi\rangle=\bigg(\widetilde{\omega}_{\zeta}+2E_{1}(\textbf{m})\bigg)C_{l_{x1}}|\psi\rangle.
\end{aligned}
\end{equation}
A similar relation holds for $C_{l_{y}}|\psi\rangle$. It should be noticed that $C_{l_{x}}|C_{l_{y}}|\psi\rangle$ is not a ground state, since it contains a flippable plaquette. Based on these observations, we can generate $2^{L_x}$ different ground states from a single reference state by applying different $C_{l_x}$ and their combinations. $2^{L_x}$ arises from $\sum_{n=0}^{L_x}C^n_{L_x}$, where $C^n_{L_x}=\frac{L_{x}!}{n!(L_{x}-n)!}$. The ground states generating by this method \cite{Shannon2004prb} are shown in Fig.\ref{colrela}. Their total number is $2\cdot 2^{L_x}+2(2^{L_y}-2)$. Therefore, the ground state degeneracy of the columnar phase is at least $2\cdot 2^{L_x}+2(2^{L_y}-2)$.

\begin{figure}[htbp]
	\centering
	\includegraphics[width=0.5\textwidth]{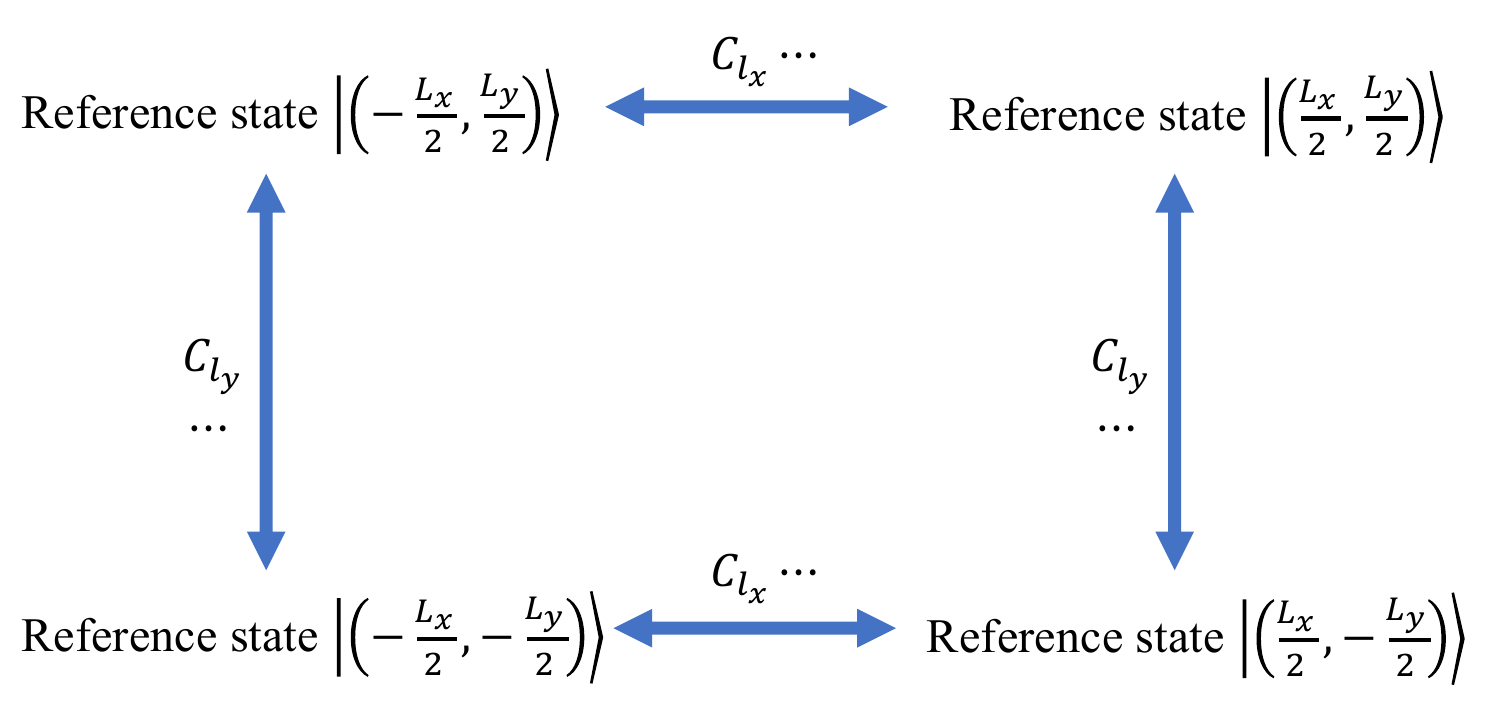}
	\caption{\label{colrela} Ground states of the columnar phase can be generated from reference states by applying various combinations of $C_{l_k}$ operators.}
\end{figure}
\bibliography{QLM_anomaly}
\end{document}